\documentclass{article}
\usepackage{spconf,amsmath,graphicx,amssymb}
\usepackage{booktabs}
\usepackage{array}
\usepackage{subfigure}
\usepackage{siunitx}

\newcommand{\R}{\mathbb{R}}
\usepackage{enumitem}
\setlist{nosep, leftmargin=14pt}

\usepackage{mwe} 

\title{Ensemble Learning for Microbubble Localization in Super-Resolution Ultrasound}
\name{Sepideh K. Gharamaleki$^{1}$ \qquad Brandon Helfield$^{2}$ \qquad Hassan Rivaz$^{1}$}

\address{$^{1}$ {Concordia University, 
	Electrical and Computer Engineering Department,
	 Montréal, QC, Canada}\\
    $^{2}${Concordia University,
	Physics and Biology Departments,
	Montréal, QC, Canada}
 }

\begin{document}
%
\maketitle
\begin{abstract}
Super-resolution ultrasound (SR-US) is a powerful imaging technique for capturing microvasculature and blood flow at high spatial resolution. However, accurate microbubble (MB) localization remains a key challenge, as errors in localization can propagate through subsequent stages of the super-resolution process, affecting overall performance. In this paper, we explore the potential of ensemble learning techniques to enhance MB localization by increasing detection sensitivity and reducing false positives. Our study evaluates the effectiveness of ensemble methods on both in vivo and simulated outputs of a Deformable DEtection TRansformer (Deformable DETR) network. As a result of our study, we are able to demonstrate the advantages of these ensemble approaches by showing improved precision and recall in MB detection and offering insights into their application in SR-US. This work was accepted in the ISBI 2025 conference. 
\end{abstract}
\section{Introduction}
\label{sec:intro}

Traditional ultrasound techniques encounter an inherent compromise between image clarity and penetration. Enhancing resolution through increasing frequency of the propagated waves, comes at the cost of increased tissue absorption, limiting the depth of effective imaging. Conversely, lower frequencies allow for deeper penetration but yield less detailed images \cite{Swanevelder2011}.

Following the break-throughs in optical super-resolution imaging \cite{Hess2006}, Ultrasound Localization Microscopy (ULM) has been proposed, which employing the strong backscatter echo properties of ultrasound contrast agent microbubbles (MBs) can, theoretically, provide a ten-fold improvement in ultrasound blood flow imaging \cite{Couture2018}. This advancement has significantly enhanced our understanding of disease states and progression, particularly in functional brain imaging \cite{Renaudin2022}, cancer \cite{Porte2024} and diabetes \cite{Zhang2023}. 

To create super-resolution ultrasound (SR-US) images of the vascular maps, MBs injected into the bloodstream are identified, localized, and frequently monitored. The MBs, confined to blood vessels, enable detailed visualization of vascular structures \cite{ChristensenJeffries2020}. 

The precision of MB localization in SR-US faces several challenges, such as Point Spread Functions (PSFs) distortion caused by MB vicinity, near-field imaging effects, variations in tissue speed of sound and attenuation \cite{Voulgaridou2025,Ashik2022}. These factors collectively contribute to alterations in PSF shape, complicating accurate localization. On the other hand, while higher MB concentrations can reduce acquisition time, they may also lead to reduced localization precision as the MB signals coincide \cite{Hingot2019}.

Addressing these issues requires careful consideration of imaging parameters, processing techniques, and potentially the development of novel algorithms that can account for these various factors to maintain localization accuracy across diverse imaging conditions.

In the past few years, several methods of MB localization have been developed and evaluated. Conventional localization methods rely on the assumption of isolated scatterers within the region of interest \cite{Heiles2022}. However, this assumption breaks down as MB trajectories intersect.

To tackle the problem of overlapping PSFs while allowing for higher MB concentrations and shorter acquisition times, various Deep Learning (DL) networks were introduced which rely on learning complex patterns from simulations and incorporating temporal context into the localization process.
A common framework employed in many DL methods is the Convolutional Neural Network (CNN) with an encoder-decoder architecture with variations in the upsampling techniques and the specific building blocks employed for feature encoding \cite{vanSloun2019,Chen2022,Milecki2021}.
Shin et al. \cite{Shin2024} introduced LOCA-ULM, which enhances localization accuracy by incorporating temporal context from adjacent frames employing two U-nets in sequence.
Building on this concept, Pustovalov et al. \cite{Pustovalov2024} recently proposed an architecture inspired by DECODE \cite{Speiser2021}. Their method utilizes 3D CNNs to incorporate temporal context, potentially allowing for the integration of information from a larger number of frames.

Transformer-based architectures have gained popularity in ULM due to their effectiveness in capturing long-range relationships in images. The ULM-TransUnet framework \cite{Zhang2024} combines the strengths of U-Net architectures and Transformer models, enhancing performance through multi-scale feature integration and long-term contextual dependencies.

Previously, we employed a DEformable DEtection TRansformer (DEDETR) solution for MB detection to better handle PSF deformations and variations in appearance, resulting in improved detection performance \cite{Gharamaleki2022, Gharamaleki2023}. 

Nevertheless, the effectiveness of any model in real-world applications is heavily based on meeting specific criteria related to data quality and hyper-parameter optimization. In these situations, ensemble learning can offer a valuable solution by combining multiple weaker models to achieve desired results \cite{Zhang2012, Casado-García2020}. 
By leveraging the strengths of multiple models, ensembles can often achieve performance levels that surpass those of individual models while manifesting increased robustness and generalization capabilities.

Beyond traditional ensemble methods, researchers have developed sophisticated techniques to refine the outputs of detection models by eliminating redundant bounding boxes. These approaches include Non-Maximum Suppression (NMS), Soft-NMS \cite{Bodla2017SoftNMS}, NMW (Non-Maximum Weighting) \cite{Ning2017NMW}, and Weighted Box Fusion (WBF) \cite{Solovyev2019WBF}. Each of these methods offers unique strategies for consolidating overlapping detections and improving overall detection accuracy.

While combining multiple machine learning models is a widespread technique for enhancing overall performance across various applications, its implementation for ULM applications presents unique challenges. For one, most of the aforementioned techniques do not detect MBs as bounding boxes and rather focus on finding the centroid of the MBs. Secondly, choosing models that while capturing a variety of different shapes and sizes of MBs in different densities, also don't overlap and repeat themselves. 
In this paper, we:

1. Propose a versatile ensemble framework for MB detection based on a transformer-based solution, \textit{i.e.} DEDETR. This approach integrates multiple detectors to leverage their complementary strengths and to optimize the ensemble's decision-making process, enhancing both the accuracy and robustness of our MB detection results.

2. Perform extensive empirical evaluation of the proposed ensemble framework and its associated ensemble strategies. Through experimentation across both simulation and in vivo datasets, we demonstrate notable improvement in the details of the generated super-resolution maps, including improved precision, recall, and RMSE scores. 

3. Analyze the framework's ability to mitigate individual detector weaknesses and adapt to diverse object scales and appearances, showcasing its effectiveness, generalizability, and robustness.

\begin{figure*}[htb]
    \centering
    \subfigure[NMS]{\includegraphics[width=0.15\textwidth, height=0.2\textheight]{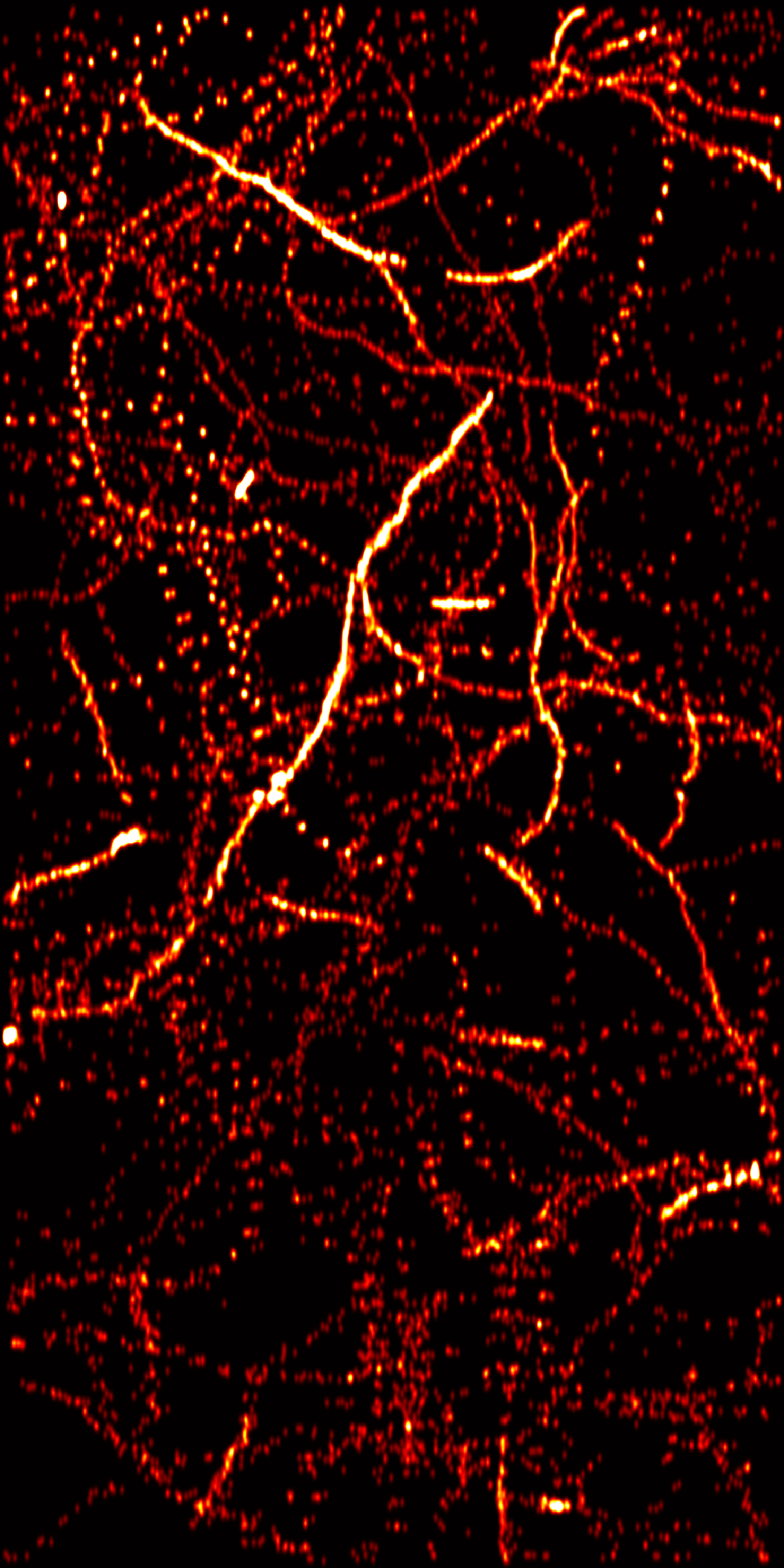}} 
    \hspace{0.1cm}
    \subfigure[NMSW]{\includegraphics[width=0.15\textwidth, height=0.2\textheight]{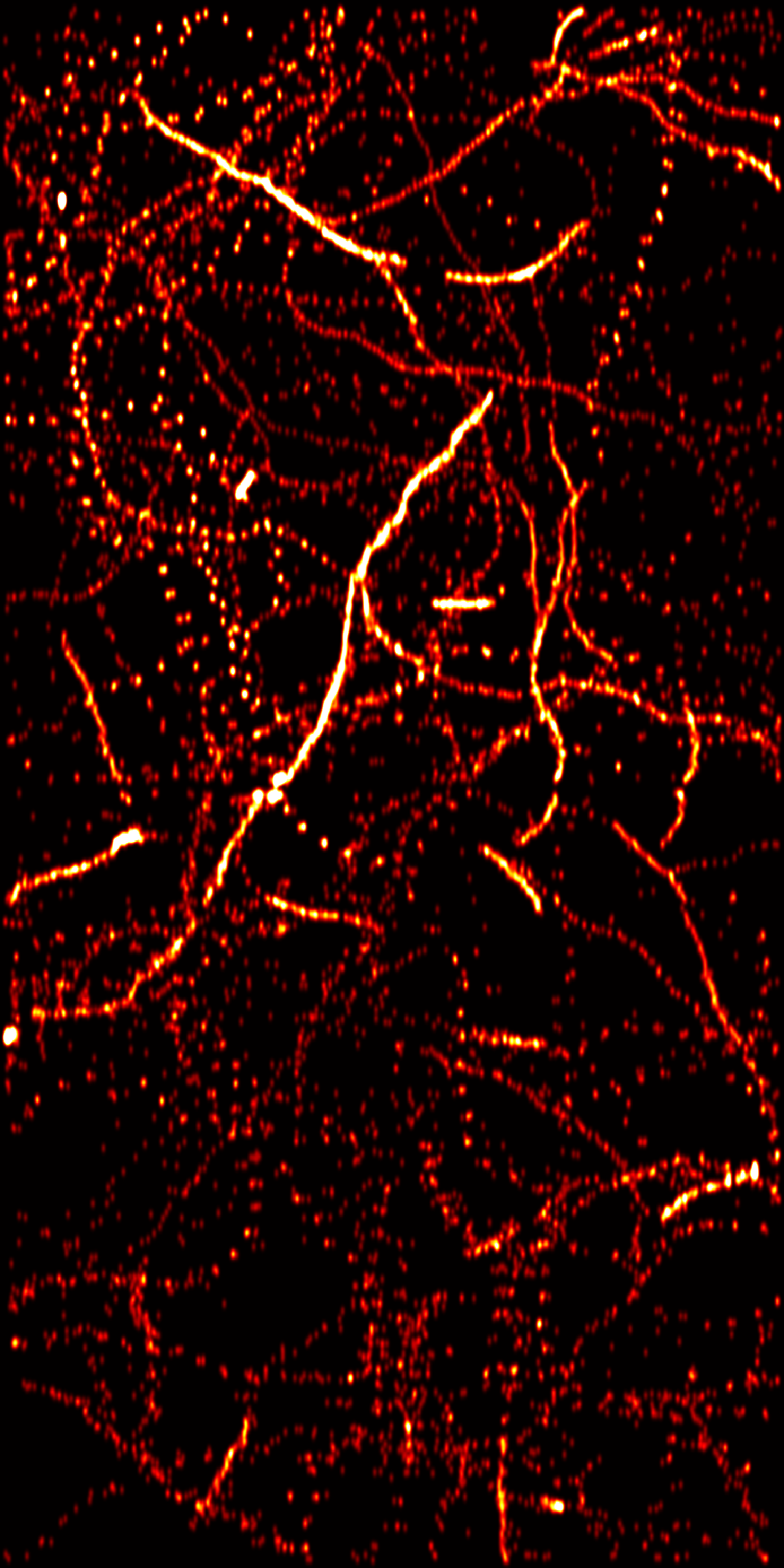}} 
    \hspace{0.1cm}
    \subfigure[Soft NMS]{\includegraphics[width=0.15\textwidth, height=0.2\textheight]{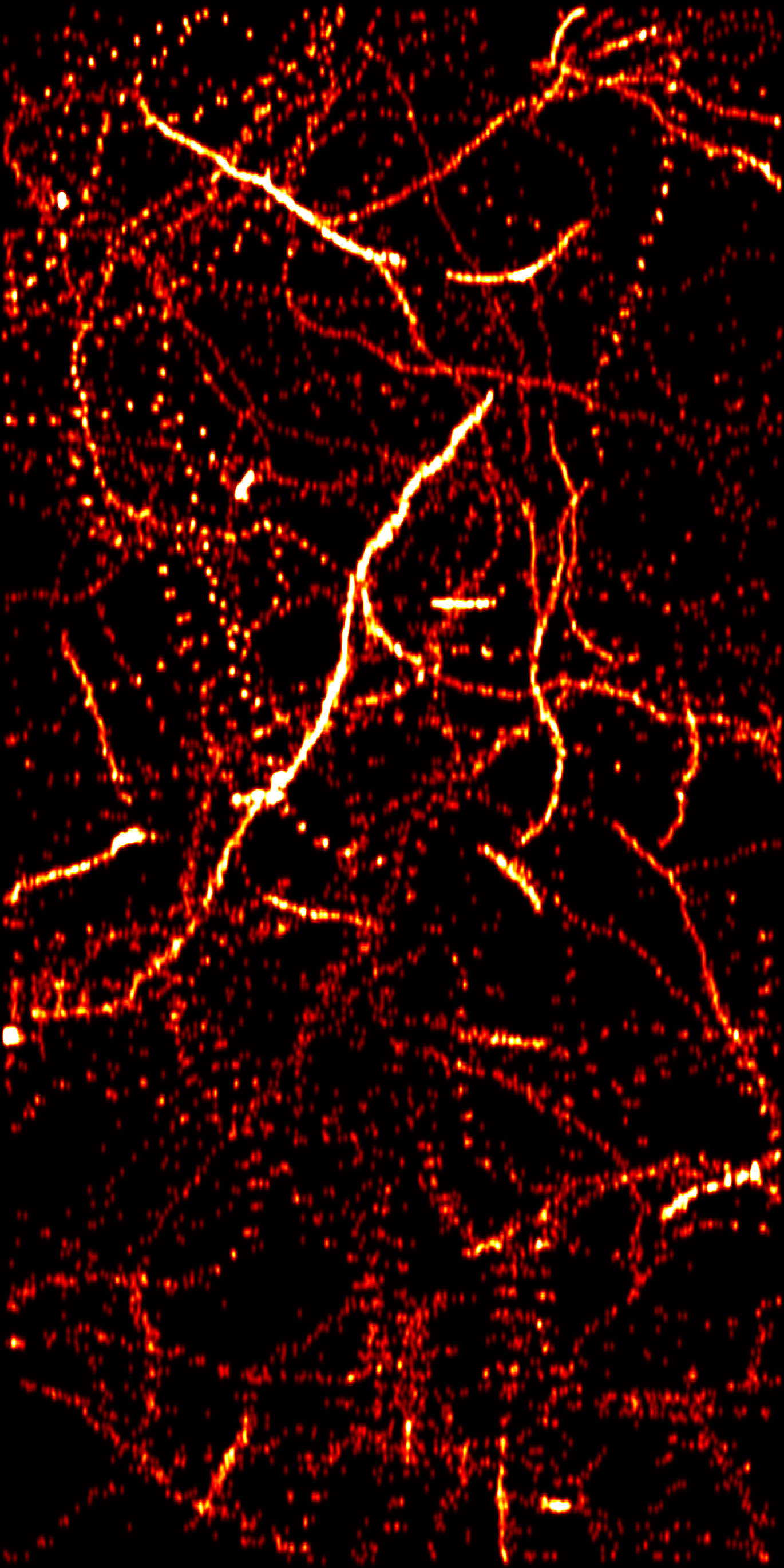}} 
    \hspace{0.1cm}
    \subfigure[WBF]{\includegraphics[width=0.15\textwidth, height=0.2\textheight]{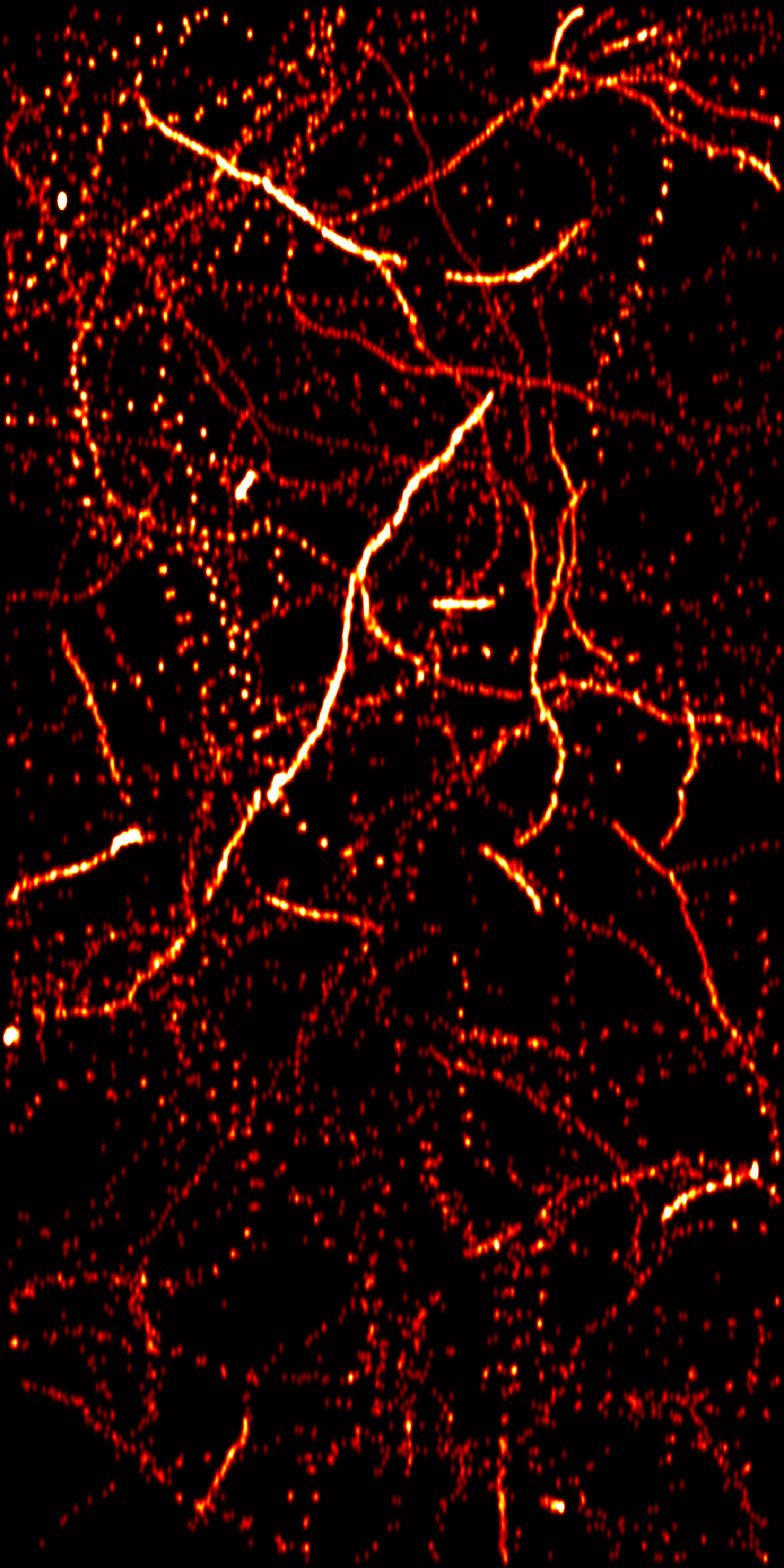}} 
    \hspace{0.1cm}
    \subfigure[DEDETR]{\includegraphics[width=0.15\textwidth, height=0.2\textheight]{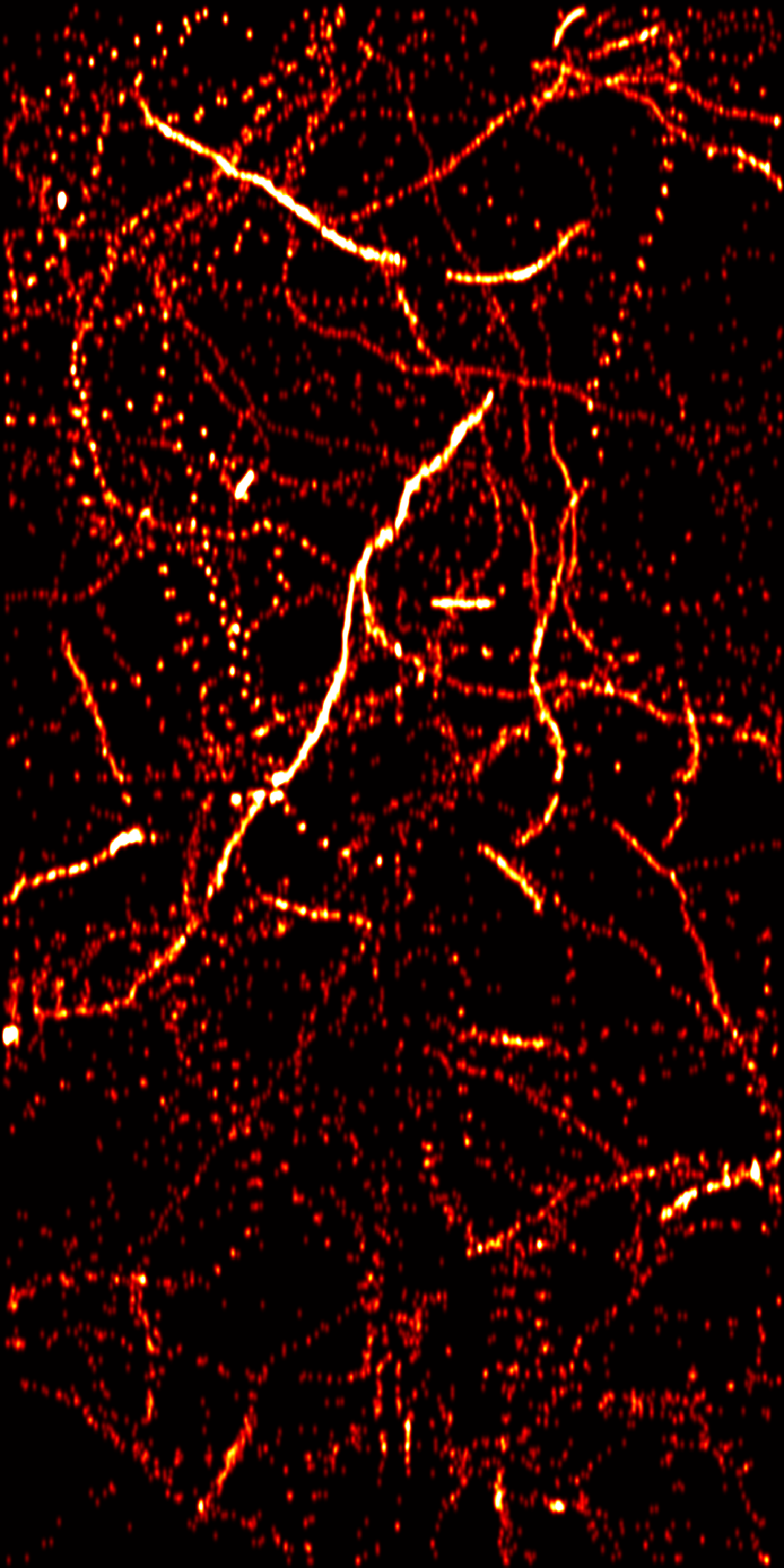}} 
    \hspace{0.1cm}
    \subfigure[GT]{\includegraphics[width=0.15\textwidth, height=0.2\textheight]{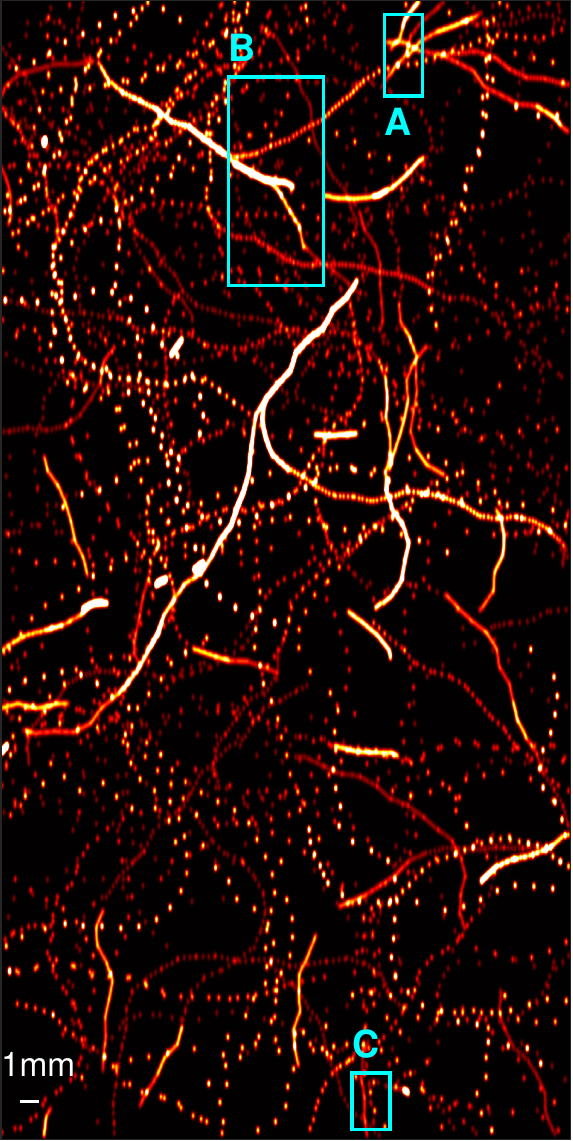}}
    
    \caption{Full-view SR maps of the simulation test dataset for different methods.}
    \label{fig:simulation}
\end{figure*}

\begin{figure}[htb]
    \centering
    (A)\\
    \includegraphics[width=0.12\textwidth,height=0.09\textheight]{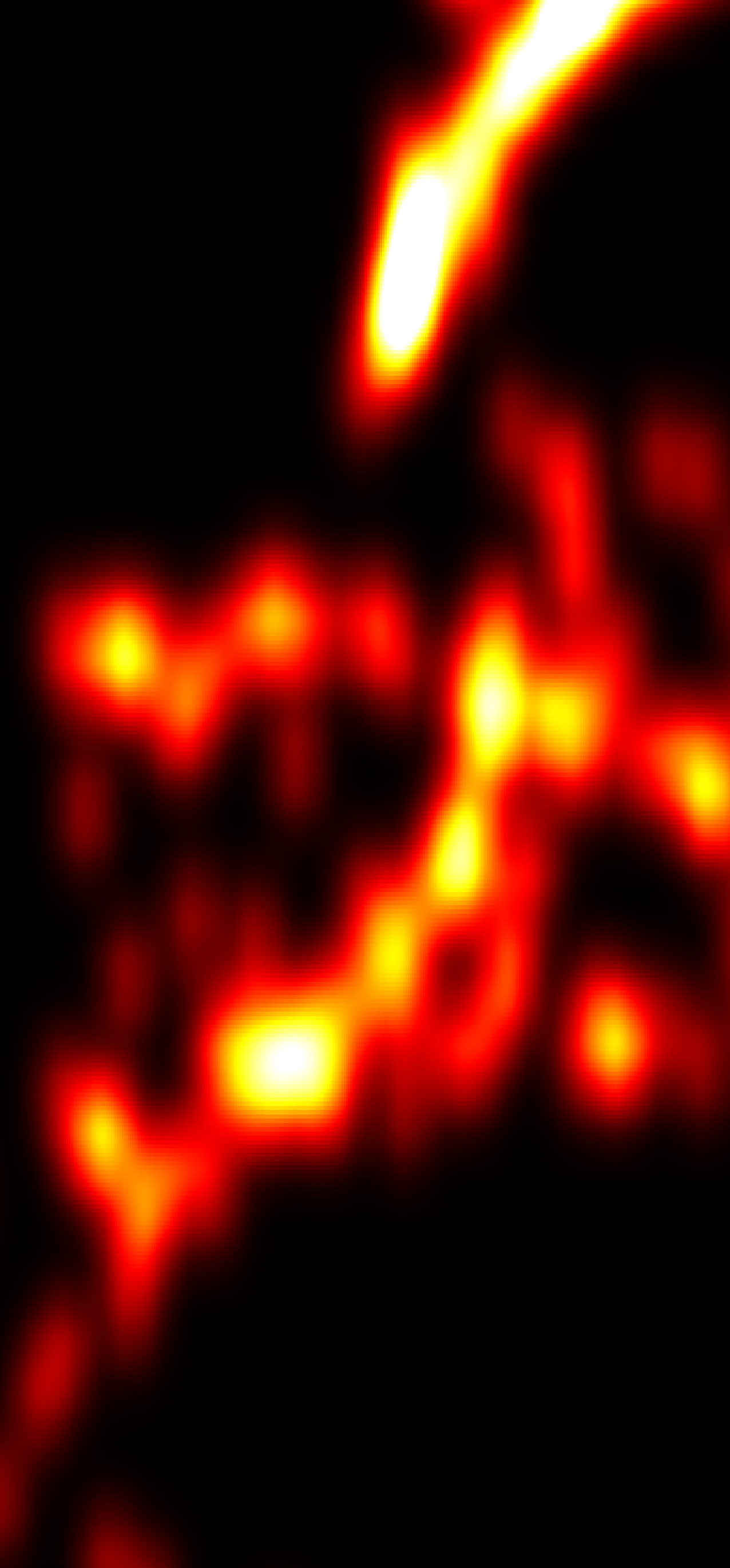}
    \hspace{0.02\textwidth}
    \includegraphics[width=0.12\textwidth,height=0.09\textheight]{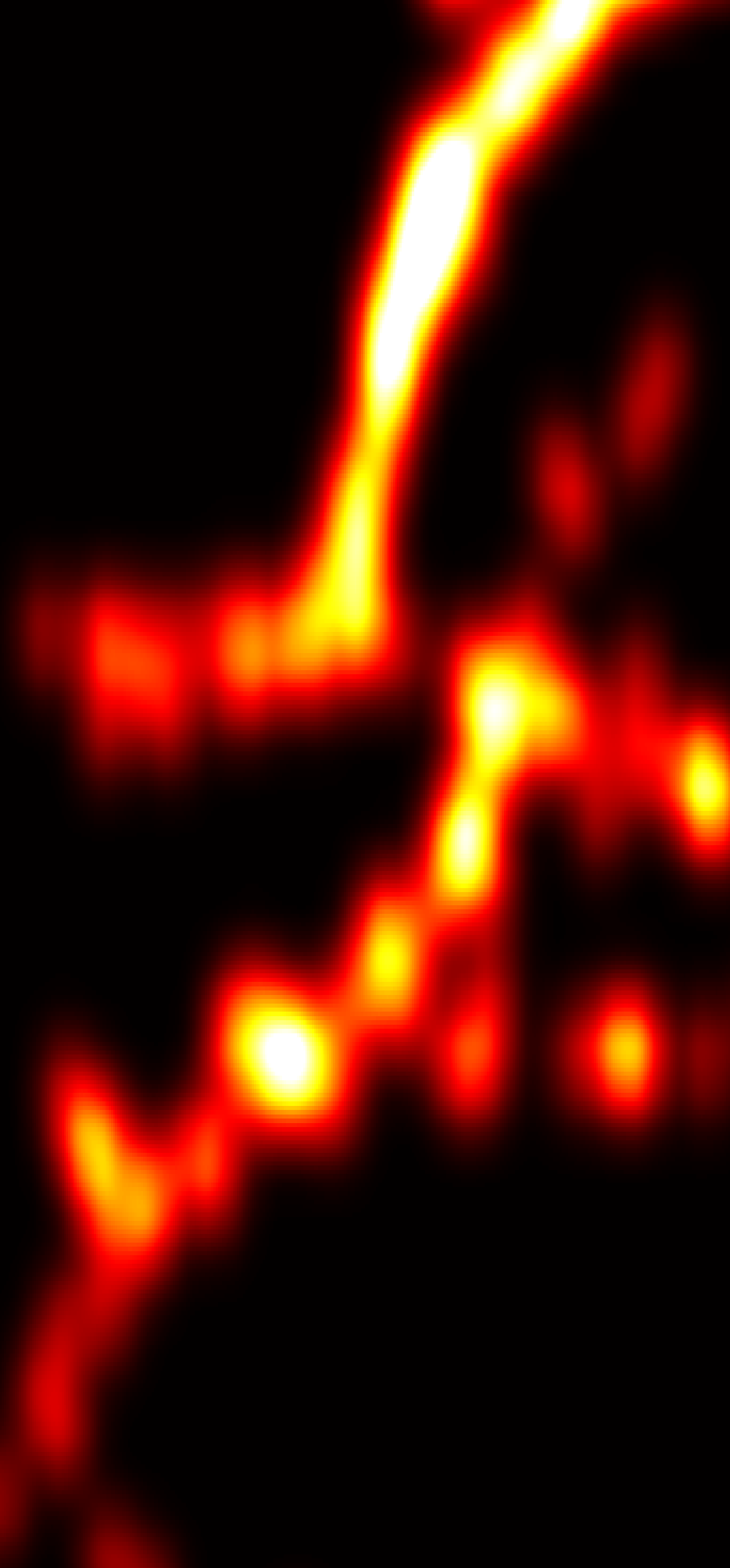} 
    \hspace{0.02\textwidth}
    \includegraphics[width=0.12\textwidth,height=0.09\textheight]{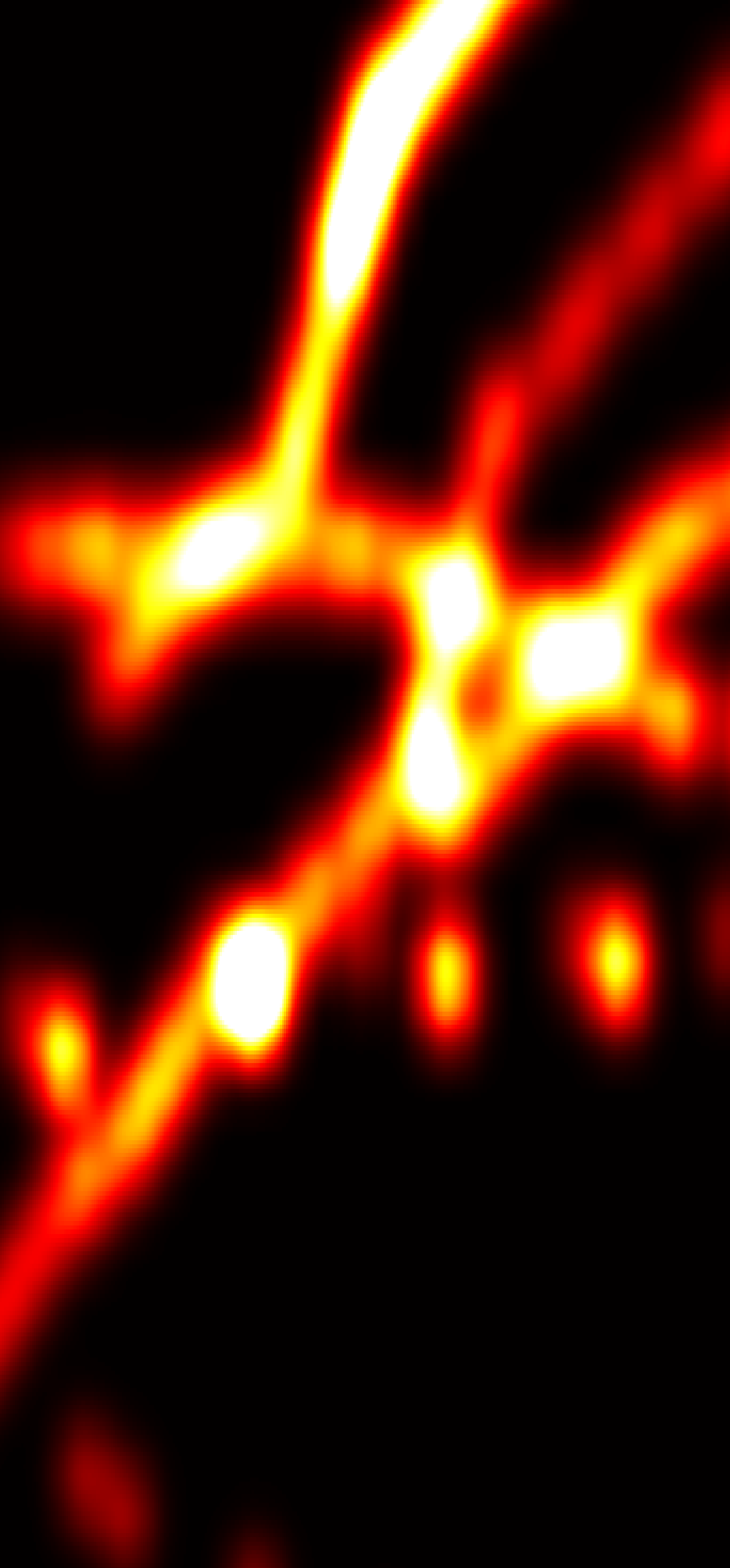}
    \\
    (B)\\
    \includegraphics[width=0.12\textwidth,height=0.09\textheight]{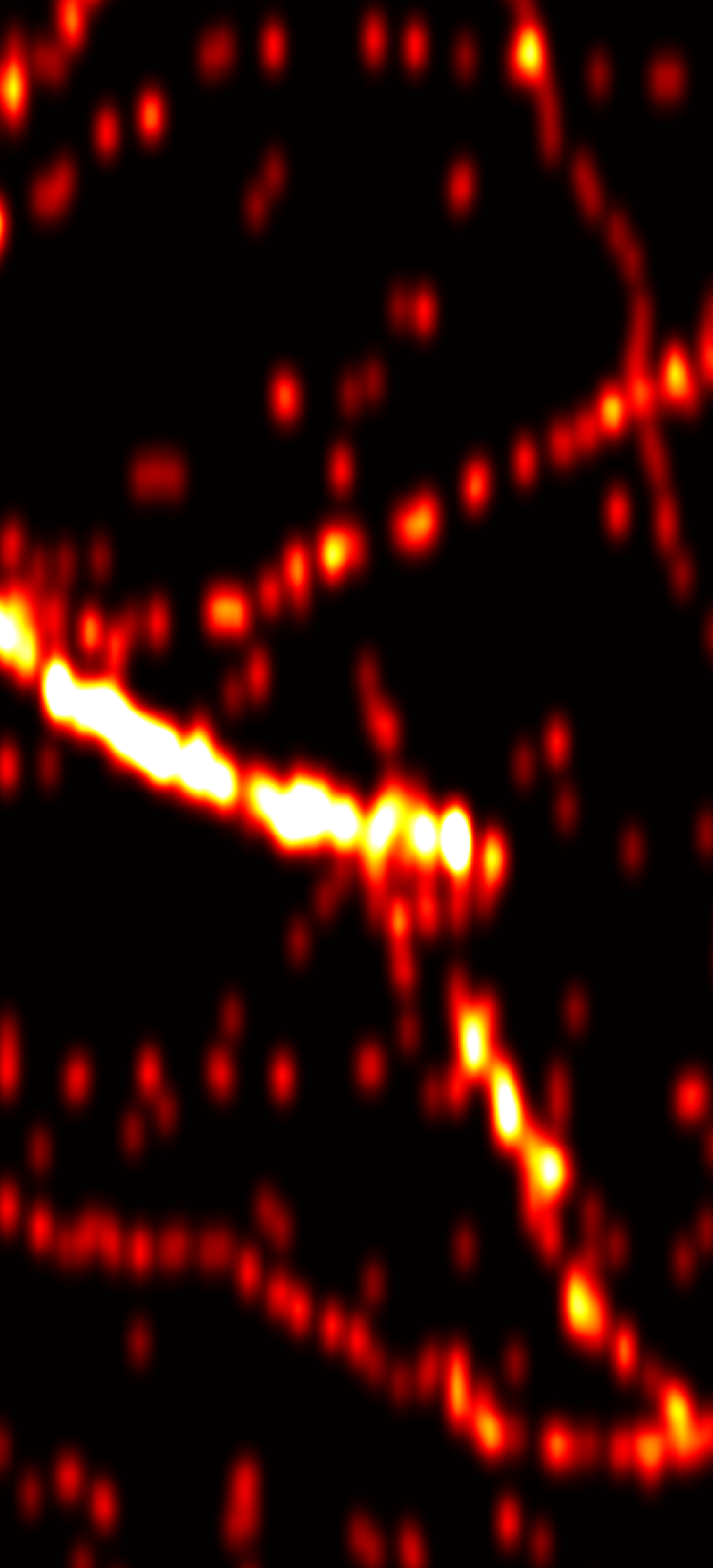} 
    \hspace{0.02\textwidth}
    \includegraphics[width=0.12\textwidth,height=0.09\textheight]{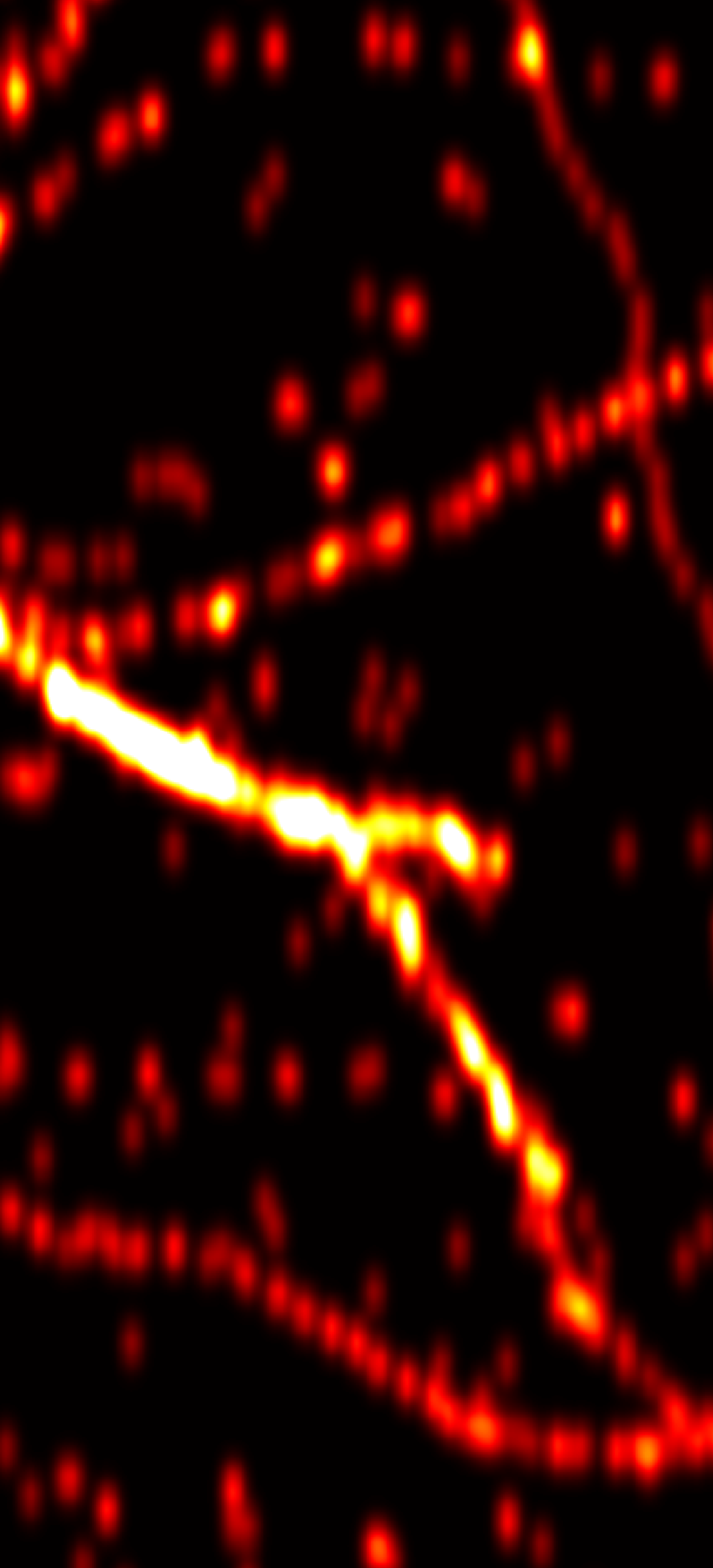} 
    \hspace{0.02\textwidth}
    \includegraphics[width=0.12\textwidth,height=0.09\textheight]{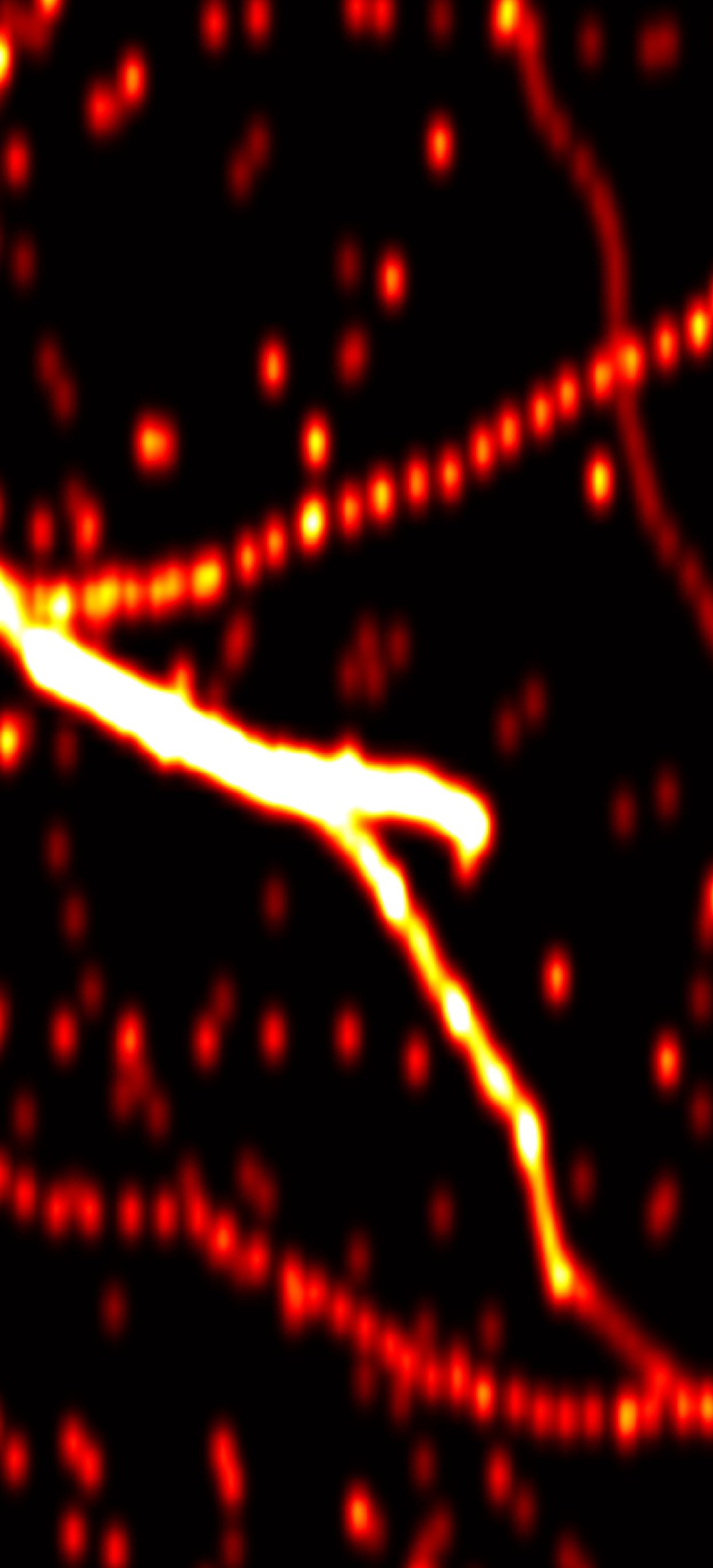}
    \\
    (C)\\
    \includegraphics[width=0.12\textwidth,height=0.09\textheight]{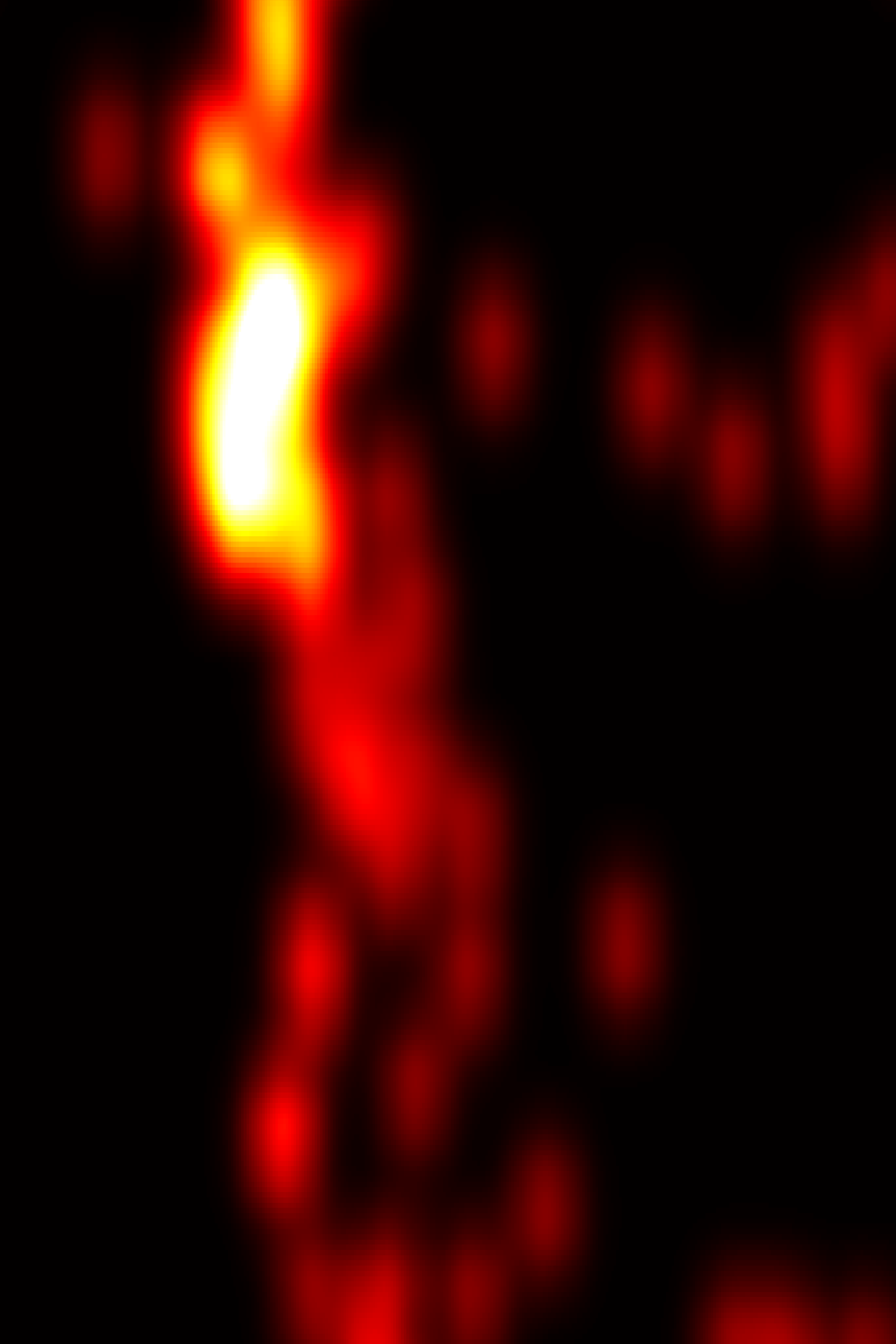} 
    \hspace{0.02\textwidth}
    \includegraphics[width=0.12\textwidth,height=0.09\textheight]{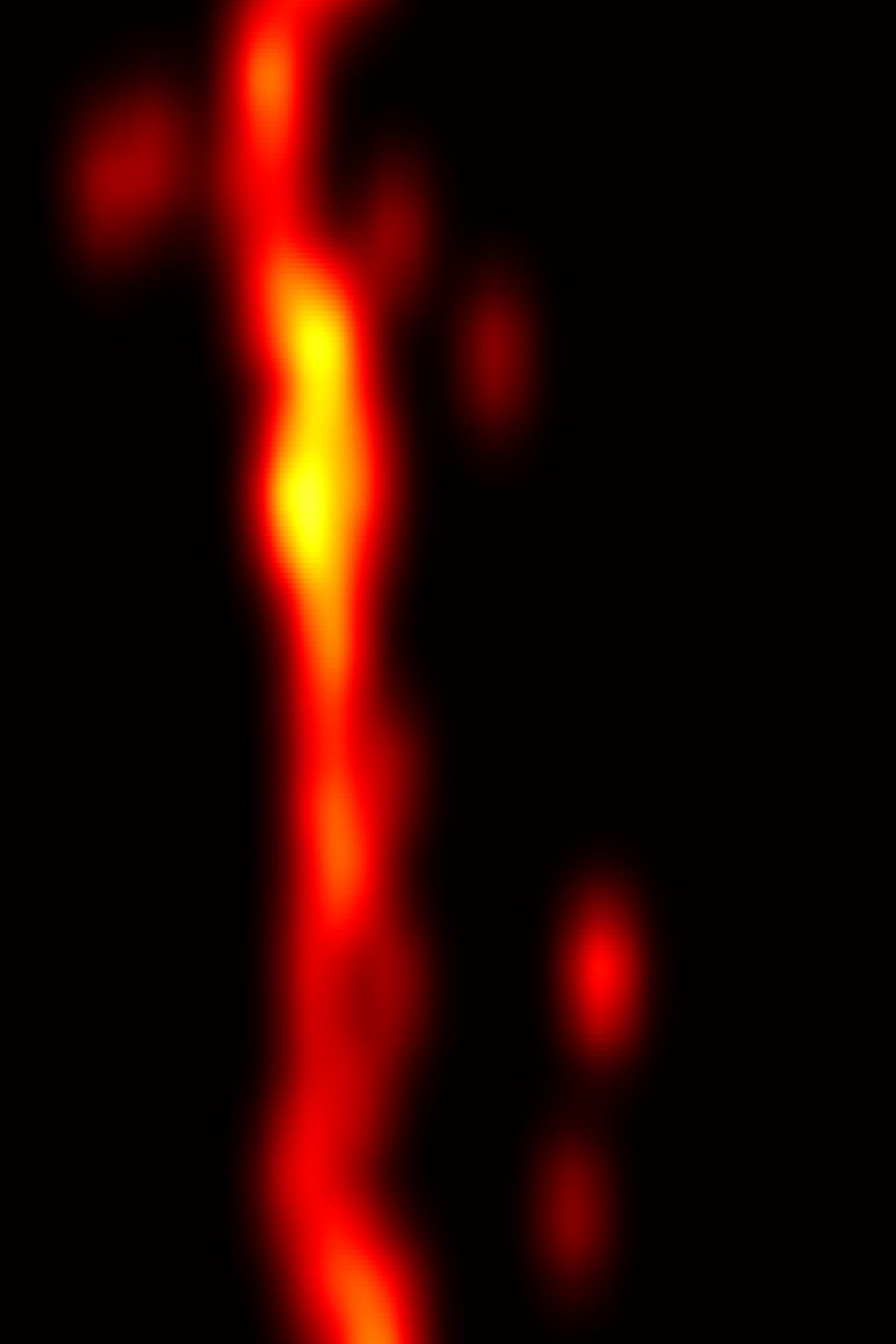} 
    \hspace{0.02\textwidth}
    \includegraphics[width=0.12\textwidth,height=0.09\textheight]{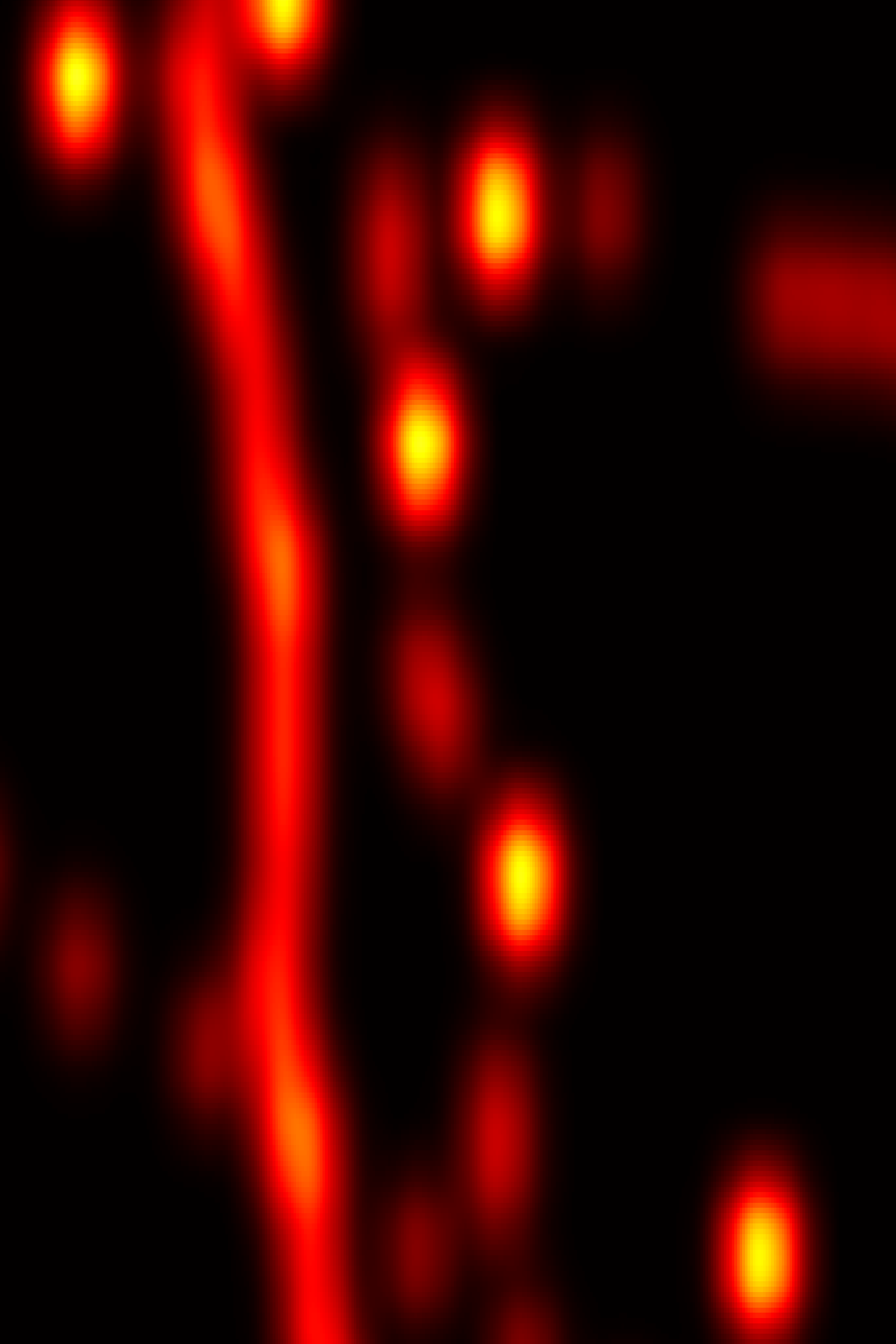}
    \\
     \begin{tabular}{p{0.12\textwidth}p{0.12\textwidth}p{0.12\textwidth}}
        \centering \makebox[0.12\textwidth]{DEDETR (Patch 4)} & 
        \centering \makebox[0.12\textwidth]{WBF} & 
        \centering \makebox[0.12\textwidth]{GT}
    \end{tabular}
    \caption{Zoomed-in boxes from different simulation SR maps, showing the results from each method (indicated by columns) for each of the (A), (B), and (C) boxes.}
    \label{fig:simboxes}
\end{figure}

\section{Methods}
\label{sec:methods}

Object detection models can be formalized as functions $h : X \mapsto Y$, where $X$ is an input image and $Y = \{y_1, \ldots, y_M\}$ is a set of detections. The formulation here is partly based on \cite{Casado-García2020}. Each detection $y_i$ is represented by a pair $(r_i, c_i)$, where:
\begin{itemize}
    \item $r_i \in \R^4$ denotes the bounding box coordinates,
    \item $c_i \in [0, 1]$ indicates the confidence score for each bounding box.
\end{itemize}

This duo is generated using our previous work, DEDETR, which was shown to improve super-resolution performance both in the sense of precision and recall.

We consider a list $LY = [Y_1,\ldots,Y_n]$, where each $Y_i$ ($i \in \{1,\ldots,n\}$) is a set of detections for a frame $X$. While $Y_i$ represents predictions from a single model.

The algorithm proceeds in four steps:

\begin{enumerate}
    \item \textbf{Prediction:} $LY$ is flattened into $E = [y_1,\ldots,y_j]$, disregarding detection model.
    
    \item \textbf{Grouping and Sorting:} Elements of $E$ are grouped based on bounding box overlap using the Intersection over Union (IoU) metric:
    
    \begin{equation}
        {IoU}(r_1, r_2) = \frac{{area}(r_1 \cap r_2)}{{area}(r_1 \cup r_2)}
    \end{equation}
    
    and then thresholded and sorted in descending order of their confidence score $c_i$.
    
    This produces $H = [YG_1,\ldots,YG_n]$, where for all 
    
    $\tilde{y} = [\tilde{r}, \tilde{c}]$, $\breve{y} = [\breve{r}, \breve{c}] \in YG_i$:
    \[IoU(\tilde{r},\breve{r}) > IOU_{thresh} \text{ and  } \tilde{c} , \breve{c} > score_{thresh}\]
    
    \item \textbf{Ensemble Strategy:} Apply one of three strategies to filter $H$:
    \begin{itemize}
        \item \textbf{NMS}: For each $YG_i$, keep only $\breve{y} = \arg\max_{y \in YG_i}(c)$, where $c$ is the detection score.
    
        \item \textbf{Soft NMS}: Update the score of each $y \in YG_i$ using:
        \begin{equation}
            c' = c \cdot g({IoU})
        \end{equation}
        where $g(\text{IoU})$ is a Gaussian or linear decay function.
    
        \item \textbf{Weighted NMS (NMSW)}: Compute the weighted average box $\bar{r}$ for each $YG_i$:
        \begin{equation}
            \bar{r} = \frac{\sum_{y \in YG_i} c \cdot r}{\sum_{y \in YG_i} c}
            \label{eq:weighed}
        \end{equation}
    
        \item \textbf{Weighted Box Fusion (WBF)}: Retains all overlapping boxes and merges them into a prediction using the same weighted average as equation \ref{eq:weighed} without suppressing any boxes. 
    \end{itemize}

    \item \textbf{Thresholding} Apply the threshold that maximizes F1-score for simulation, or generates the most visually appealing results for in vivo.
\end{enumerate}

Finally the centers of MBs are found by finding the centers of the ensemble detection list $Y = \{y_1, \ldots, y_M\}$ of ensemble detections.

\begin{figure*}[htb]
    \centering
    \subfigure[NMS]{\includegraphics[width=0.18\textwidth, height=0.2\textheight]{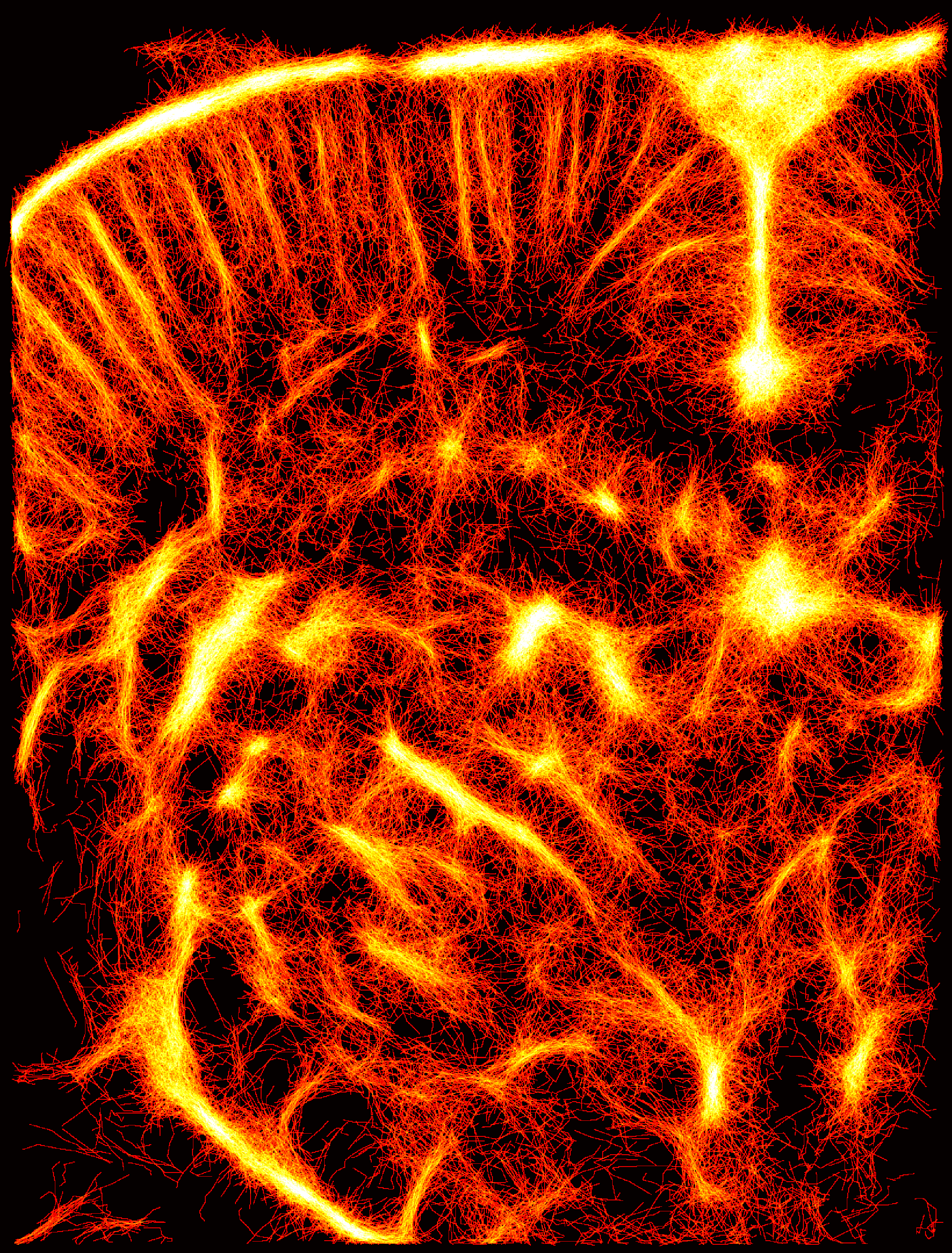}} 
    \hspace{0.01\textwidth}
    \subfigure[NMSW]{\includegraphics[width=0.18\textwidth, height=0.2\textheight]{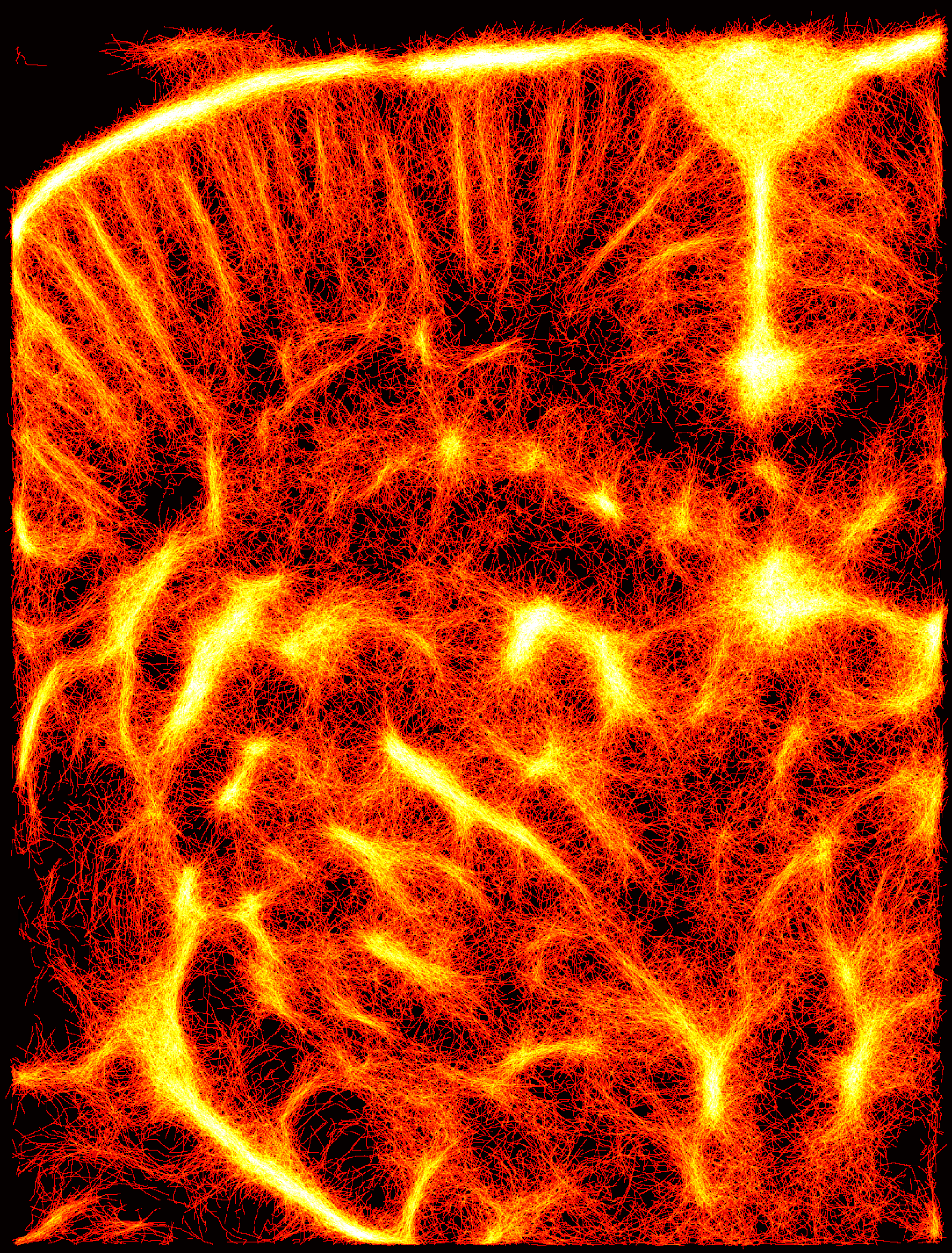}} 
    \hspace{0.01\textwidth}
    \subfigure[Soft NMS]{\includegraphics[width=0.18\textwidth, height=0.2\textheight]{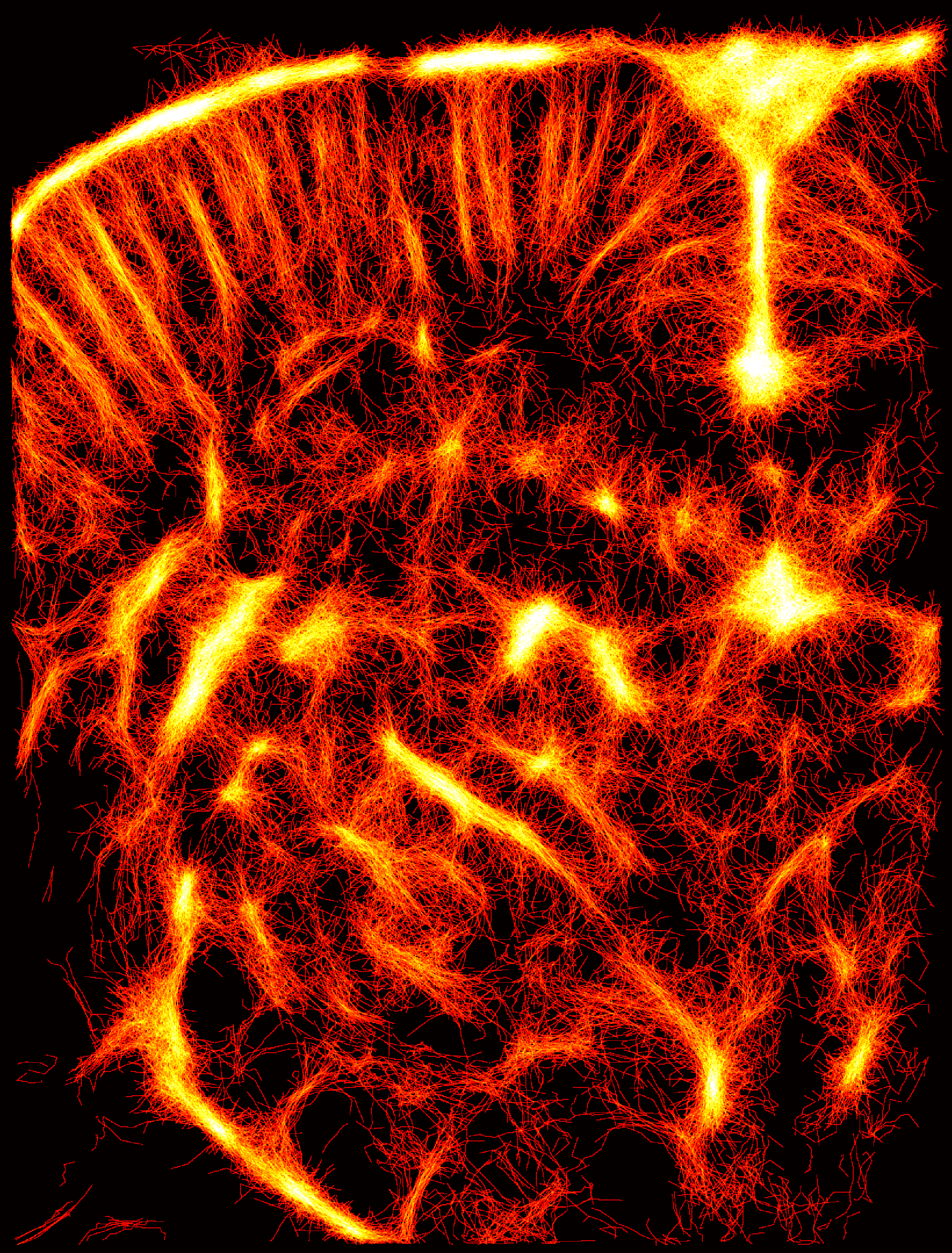}} 
    \hspace{0.01\textwidth}
    \subfigure[WBF]{\includegraphics[width=0.18\textwidth, height=0.2\textheight]{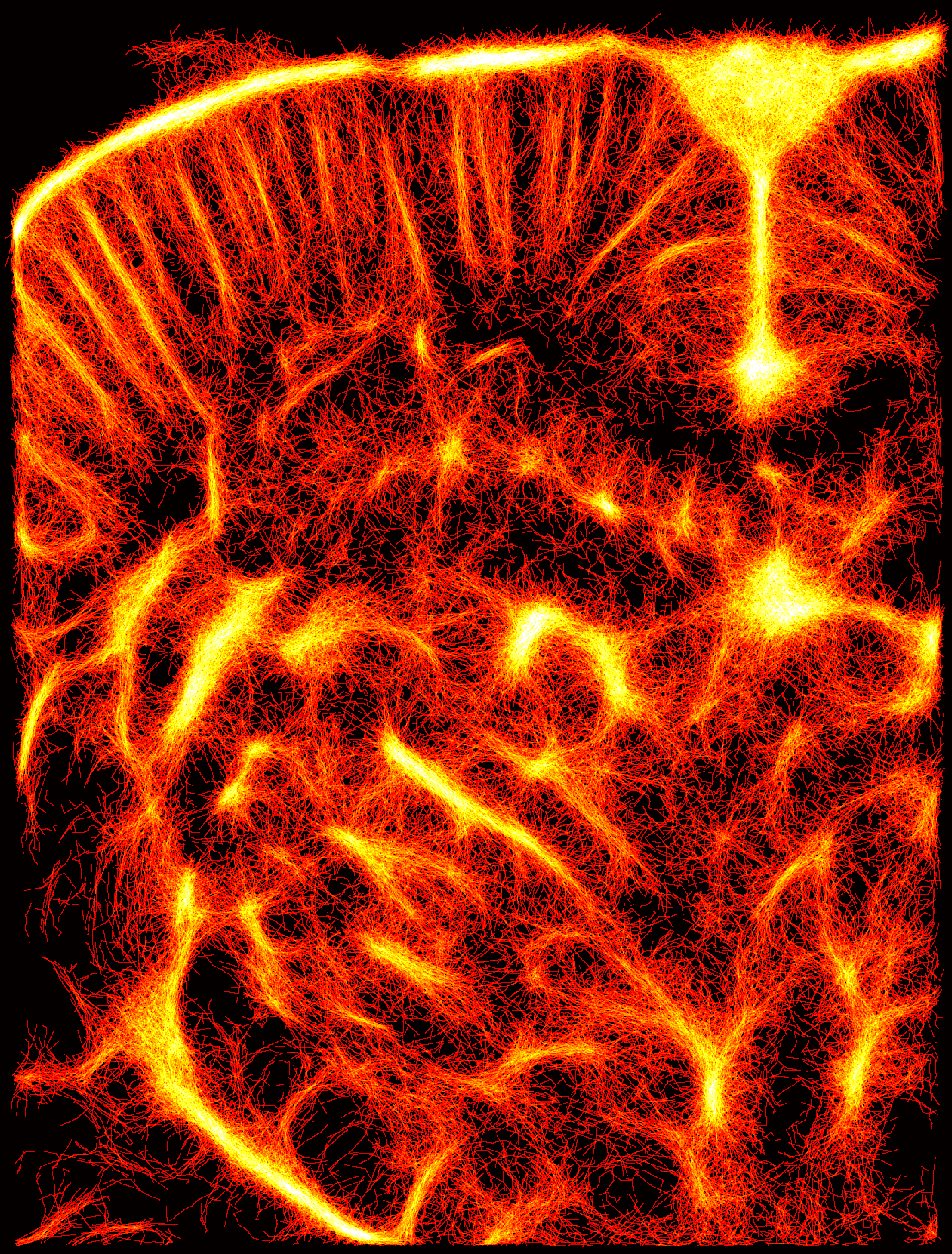}} 
    \hspace{0.01\textwidth}
    \subfigure[DEDETR]{\includegraphics[width=0.18\textwidth, height=0.2\textheight]{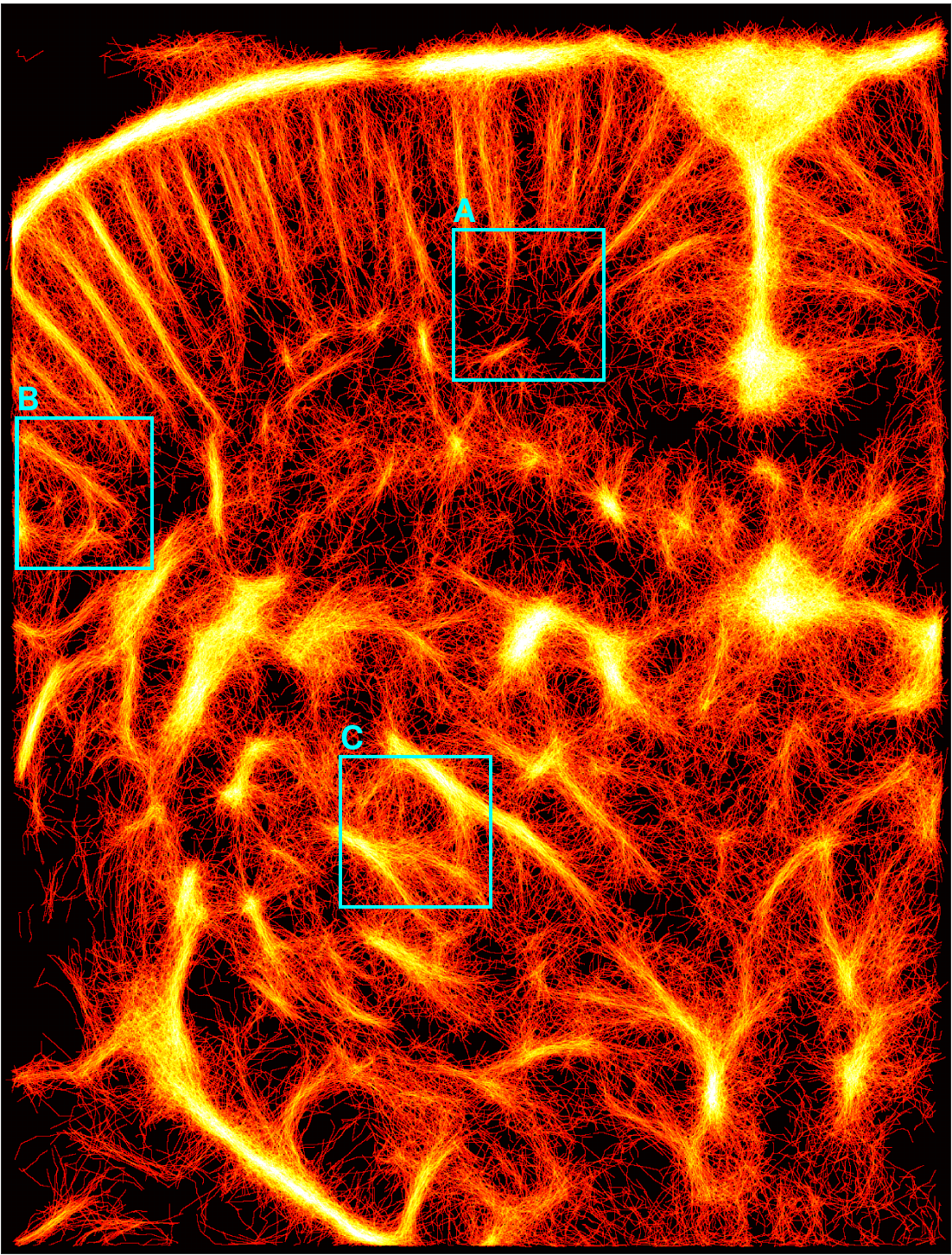}} 
    \caption{Full-view SR maps of the \textit{in vivo} test dataset for different methods.}
    \label{fig:invivo}
\end{figure*}

\begin{figure}[htb]
    \centering
    (A)\\
    \includegraphics[width=0.12\textwidth,height=0.09\textheight]{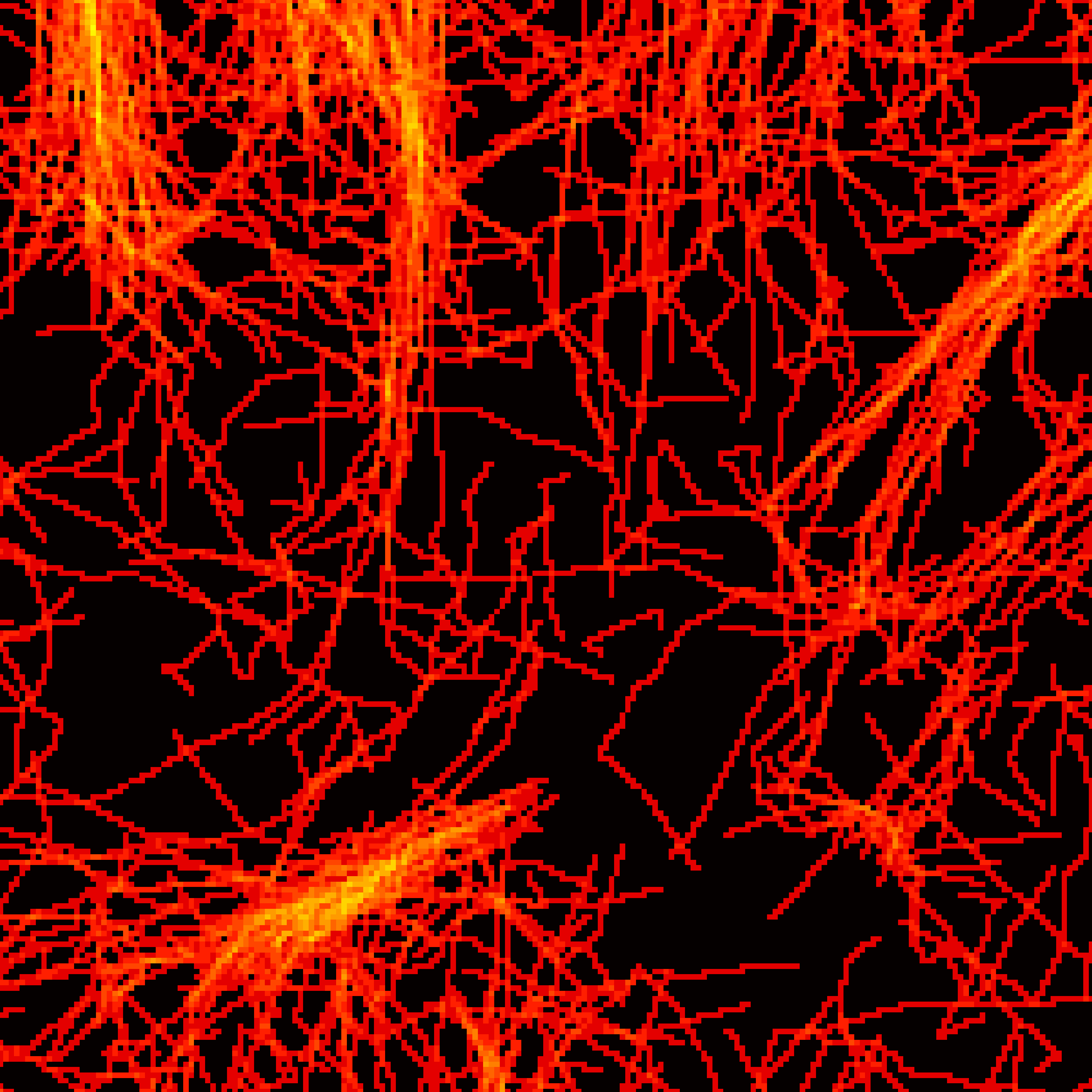}
    \hspace{0.02\textwidth}
    \includegraphics[width=0.12\textwidth,height=0.09\textheight]{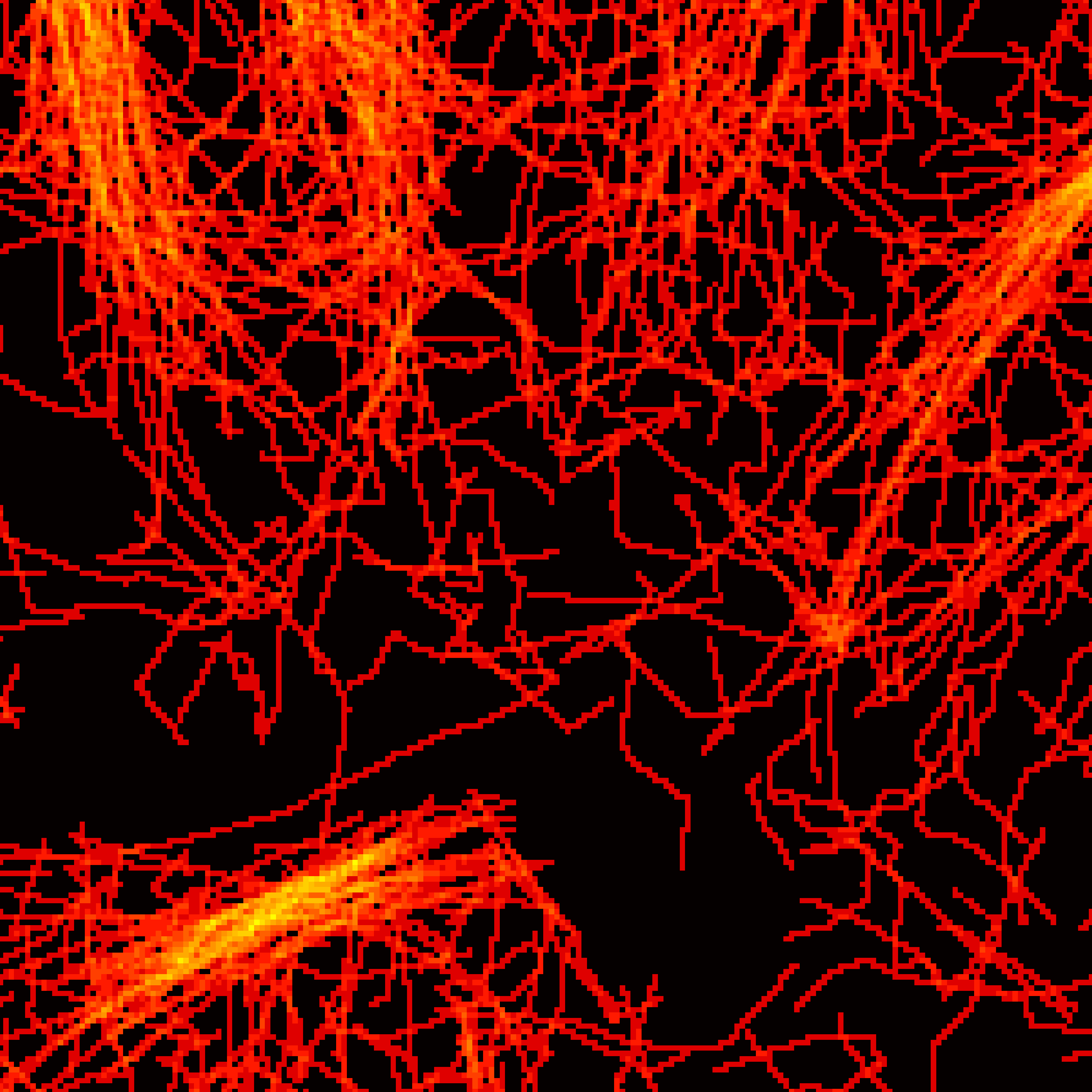}
    \\
    (B)\\
    \includegraphics[width=0.12\textwidth,height=0.09\textheight]{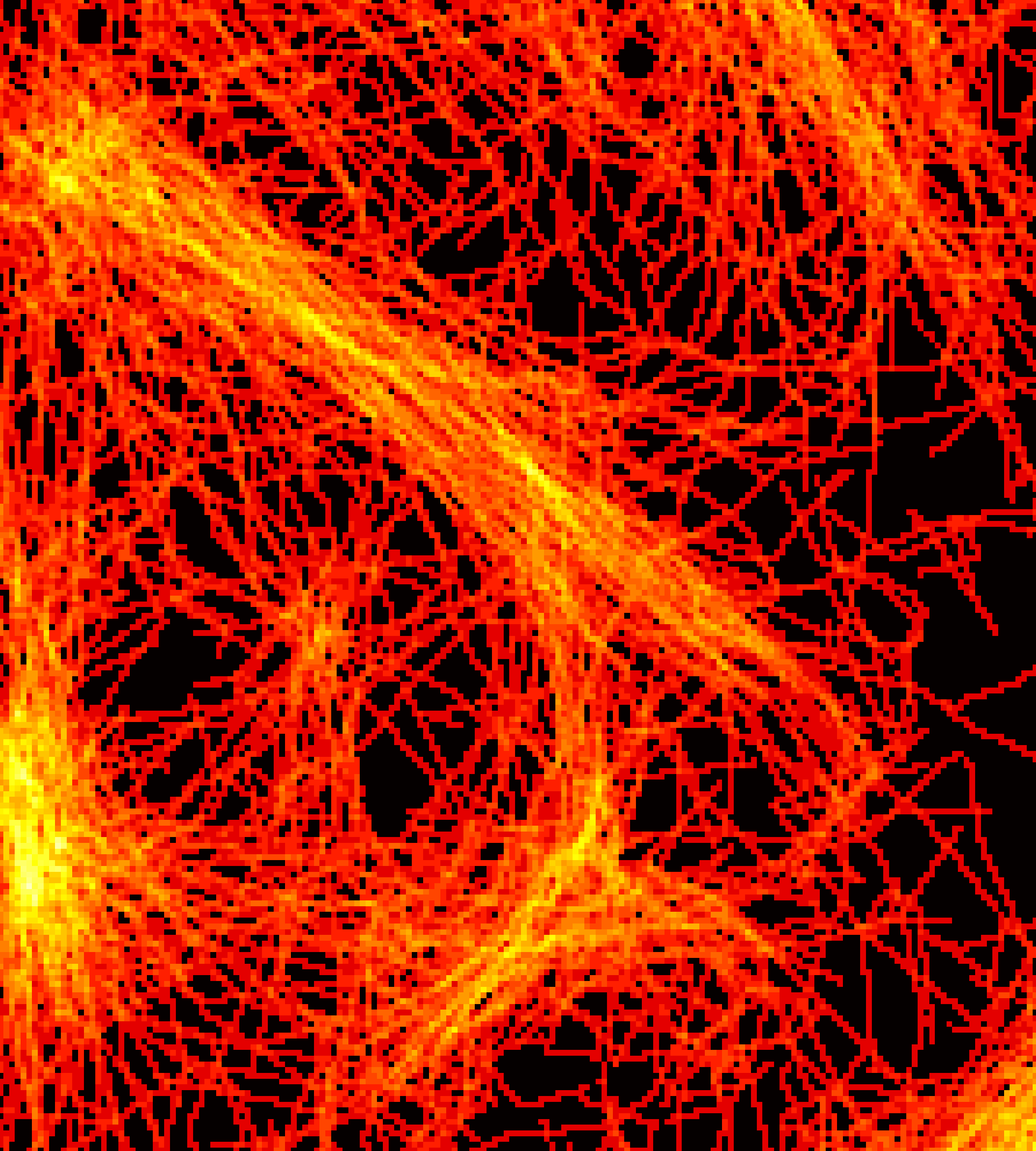}
    \hspace{0.02\textwidth}
    \includegraphics[width=0.12\textwidth,height=0.09\textheight]{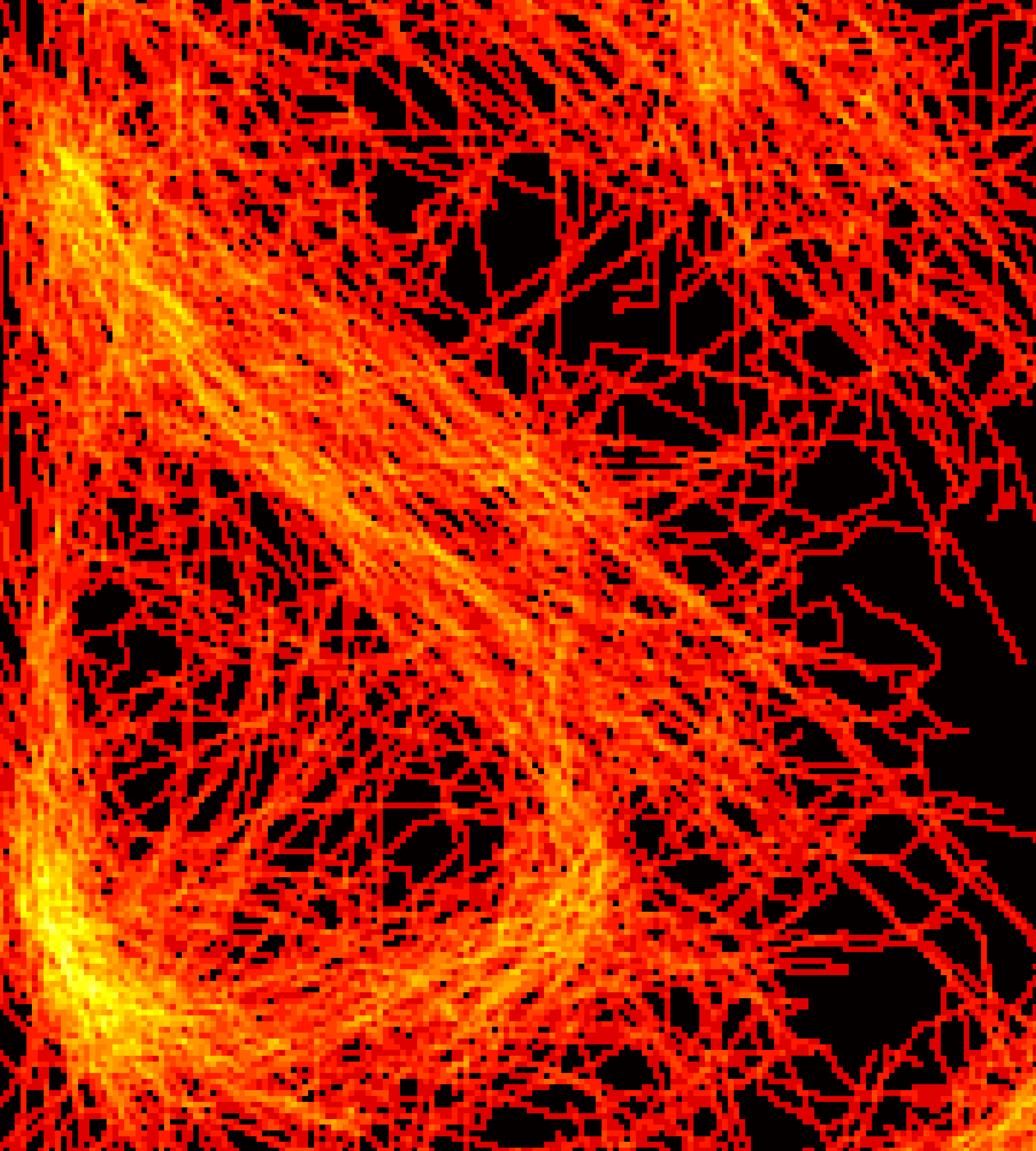}
    \\
    (C)\\
    \includegraphics[width=0.12\textwidth,height=0.09\textheight]{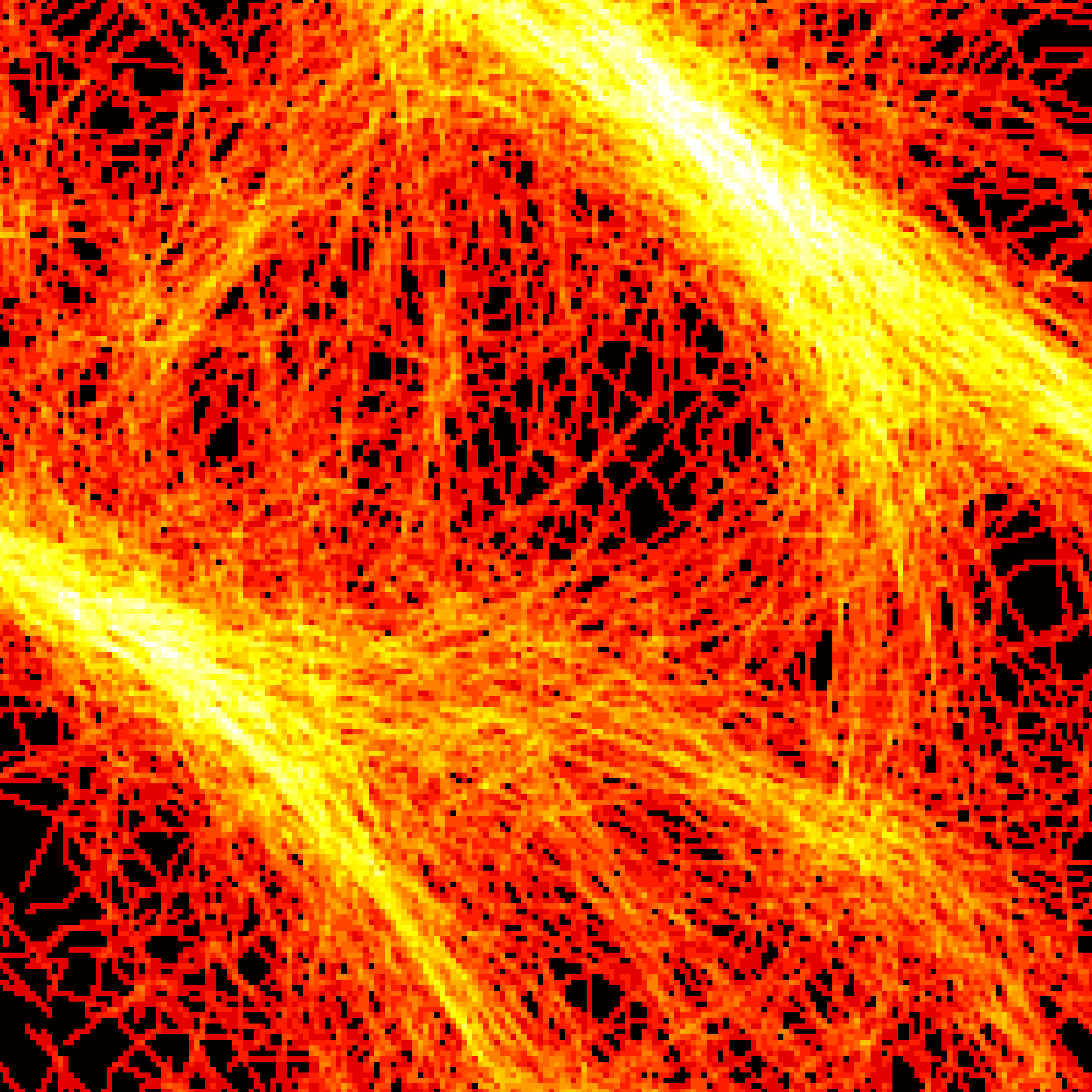}
    \hspace{0.02\textwidth}
    \includegraphics[width=0.12\textwidth,height=0.09\textheight]{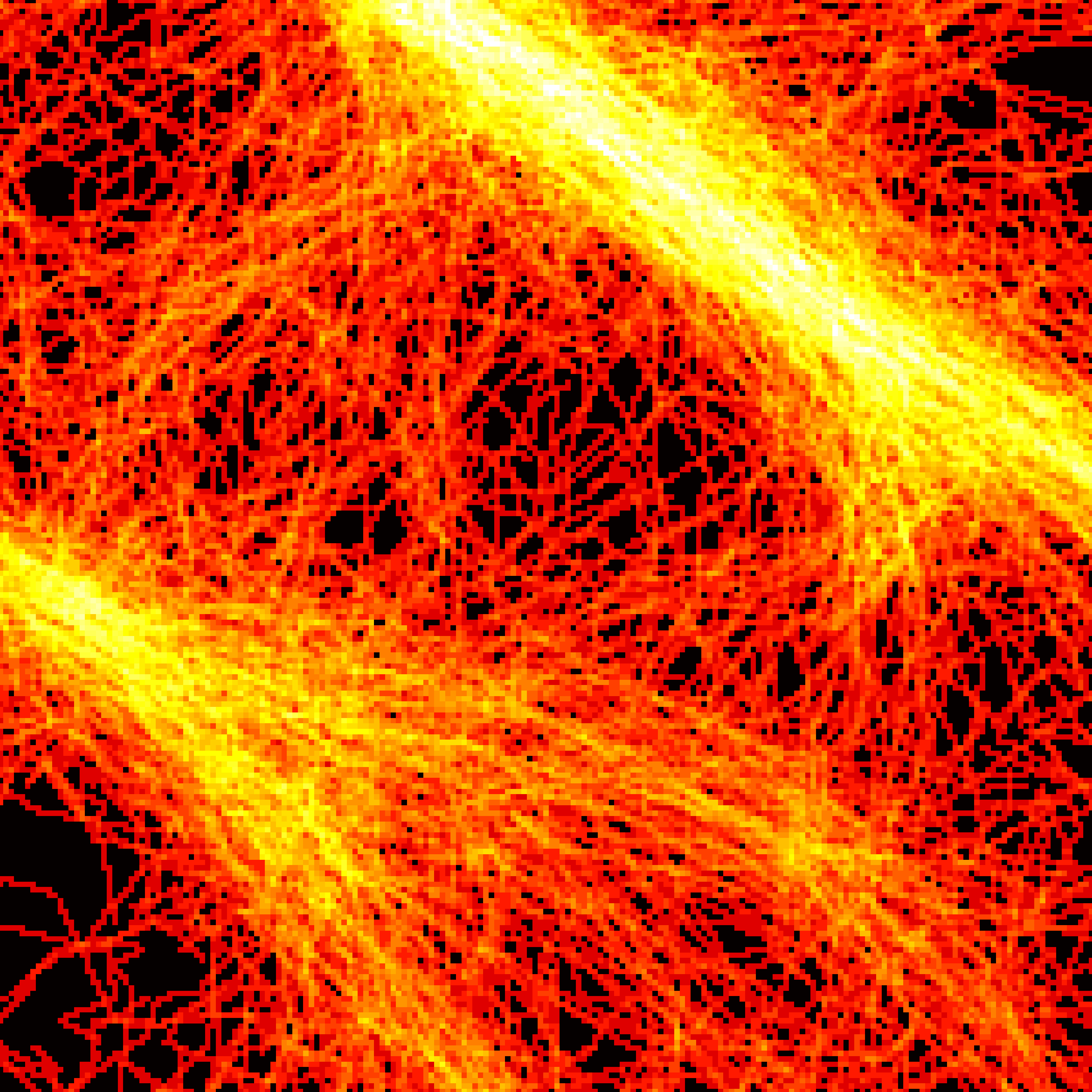}
    \\
    \makebox[0.12\textwidth]{DEDETR Patch 4}
    \hspace{0.02\textwidth}
    \makebox[0.12\textwidth]{WBF}
    
    \caption{Zoomed-in boxes from different \textit{in vivo} SR maps, showing the results from each method (indicated by columns) for each of the (A), (B) and (C) boxes.}
    \label{fig:invivoboxes}
\end{figure}

Each of the ensemble strategies for handling overlapping detections in object detection has its strengths and limitations. NMS is popular for its simplicity and efficiency, retaining only the highest-scoring box while discarding others. However, this can lead to missed detections in high bubble-density regions due to its aggressive suppression. Soft-NMS mitigates this issue by reducing the scores of overlapping boxes rather than eliminating them, preserving more detections but with higher computational cost. NMSW strikes a balance by merging overlapping boxes through confidence-weighted averaging while still suppressing lower-scoring ones, improving localization accuracy but potentially discarding valid detections. Finally, WBF takes a different approach, retaining and merging all overlapping boxes to create a more accurate final prediction, making it ideal for our application. However, WBF is computationally intensive and can increase the risk of false positives (FPs) by incorporating weaker predictions. Each method thus offers trade-offs between accuracy, computational efficiency, and the preservation of overlapping detections, depending on the task requirements.
\begin{table}
\caption{Metric comparison between models trained with different number of patches.}
\centering
\begin{tabular}{@{}l!{\vrule width 0.6pt}c!{\vrule width 0.6pt}c!{\vrule width 0.6pt}c@{}}
\specialrule{2pt}{0pt}{3pt}
Models & Precision (\%) & Recall (\%) & RMSE ($mm$) \\
\specialrule{0.6pt}{3pt}{3pt}

Patch 2 & 61.38 & 67.65 & 0.1212\\
Patch 4 & 82.92 & 71.55 & 0.1089\\
Patch 6 & 82.76 & 71.89 & 0.1108\\
Patch 8 & 86.27 & 66.60 & 0.1153\\
Patch 12 & 82.02 & 68.51 & 0.1118\\
\specialrule{3pt}{3pt}{3pt}
\end{tabular}

\label{table:patches}
\end{table}

\begin{table}
\caption{Metric comparison between models with different ensemble strategies}
\centering
\begin{tabular}{@{}l!{\vrule width 0.6pt}c!{\vrule width 0.6pt}c!{\vrule width 0.6pt}c@{}}
\specialrule{2pt}{0pt}{3pt}
Models & Precision (\%) & Recall (\%) & RMSE ($mm$) \\
\specialrule{0.6pt}{3pt}{3pt}

NMS & 87.32 & 75.03 & 0.1101\\
Soft NMS& 78.20 & 77.66 & 0.1079\\
NMSW & 89.71 & 74.55 & 0.1042\\
WBF & 93.69 & 74.33 & 0.1027\\
\specialrule{3pt}{3pt}{3pt}
\end{tabular}

\label{table:ensembles}
\end{table}
\section{Results}
\label{sec:results}

The data used for evaluating our method is from IEEE IUS 2022 Ultra-SR Challenge \cite{Lerendegui2024}. Our five networks are set as DE-DETR networks trained on different patch sizes. In this way, we can capture a variety of MB shapes in both higher density regions and deeper parts of imaging. Table \ref{table:patches} shows the metrics as a result of the inference using each network.

Using the following parameters:
\[IOU_{thresh} = 0.2 \quad \text{ ,and  } score_{thresh} = 0.01\]

and $$weights = [0.1, 0.8, 0.7, 0.6, 0.5]$$ for each the networks in Table \ref{table:patches}, respectively, ensemble strategies are implemented.


Table \ref{table:ensembles} shows the results for each ensemble strategy, which demonstrates a 11\% and 3\% increase in precision and recall and around 0.01$mm$ decrease in ovearall RMSE. 

In Figures \ref{fig:simulation} and \ref{fig:simboxes}, SR maps using the best results of our five networks (namely the network trained with 4 patches for each frame), ensemble learning results, and the zoomed-in versions are mentioned for simulation test data. It's important to mention that no tracking algorithms are applied and the SR maps are only generated using MB localization. The results demonstrate that all ensemble strategies effectively reduce FPs while also enhancing the detection of previously missed MBs. As anticipated, WBF outperforms the other ensemble methods by identifying more vessels. NMS and NMSW produce similar results, while soft NMS generates more FPs than WBF.

The same evaluation for \textit{in vivo} can be found in Figures \ref{fig:invivo} and \ref{fig:invivoboxes}. Determining the improvements in \textit{in vivo} data is more challenging due to the lack of access to ground truth. However, similar to the simulations, consistent improvements can still be observed. 
In both the full and zoomed-in comparisons, WBF captures more detailed vessel structures with fewer FPs. The original network tends to miss finer details and provides sparser representations in certain regions. NMS and NMSW underperform, detecting fewer vessels, while Soft NMS misses many MB tracks. Though harder to validate compared to the simulation dataset, WBF seems to somewhat improve the MB detection process by finding finer details.

\section{Conclusion}
\label{sec:conclusion}

Our research demonstrates that ensemble techniques can significantly enhance MB localization accuracy, potentially improving disease diagnosis. However, these benefits come with trade-offs such as longer inference times, increased computational resource demands, greater complexity in debugging and interpreting predictions.
While our experiments show that ensembling yields better results, the performance gains are not always proportional to the additional resources required. The enhanced reliability of our ensemble approach has significant implications for MB detection solutions, but implementation should be tailored to specific use cases.
Future work should focus on optimizing the balance between performance gains and resource utilization, as well as improving the interpretability of ensemble predictions in clinical settings.



\section{Compliance with ethical standards}
\label{sec:ethics}

This research study was conducted retrospectively using datasets made available by \cite{Lerendegui2024}. This is a numerical simulation study for which no ethical approval was required.

\section{Acknowledgments}
\label{sec:acknowledgments}

This work was supported by the Natural Sciences and
Engineering Research Council of Canada (NSERC).

\bibliographystyle{IEEEbib}
\begingroup
\footnotesize
\bibliography{refs}
\endgroup

\end{document}